\documentclass[iop,apj,numberedappendix,twocolumn, tighten]{aastex62}

\usepackage{arydshln}
\usepackage{natbib}
\usepackage{graphicx}
\usepackage{amsmath,amsfonts}
\usepackage{multirow}
\usepackage{xspace}
\usepackage[normalem]{ulem}
\usepackage{booktabs}
\usepackage{colortbl}
\usepackage{hyperref}
\usepackage[nameinlink]{cleveref}
\usepackage[encapsulated]{CJK}
\usepackage{ucs}
\usepackage{float}
\usepackage[utf8]{inputenc}

\usepackage{eht}
\usepackage{newtxtext}
\usepackage{tablefootnote} 
\usepackage{mathrsfs}
\usepackage{lineno}

\definecolor{fed_blue}{HTML}{07004D}
\definecolor{steel_blue}{HTML}{2D82B7}
\definecolor{steel_blue_dark}{HTML}{1C71A6}
\definecolor{aqua_marine}{HTML}{42E2B8}
\definecolor{dutch_white}{HTML}{F3DFBF}
\definecolor{light_coral}{HTML}{EB8A90}
\definecolor{light_coral_dark}{HTML}{BA5A60}

\hypersetup{
    colorlinks = true,
    linkcolor = purple,
    urlcolor  = steel_blue_dark,
    citecolor = steel_blue_dark,
    anchorcolor = black
}

\graphicspath{{./figures}}

\def\lsim{\mathrel{\raise.3ex\h box{$<$\kern-.75em\lower1ex\hbox{$\sim$}}}}
\def\gsim{\mathrel{\raise.3ex\hbox{$>$\kern-.75em\lower1ex\hbox{$\sim$}}}}
\def\gtwid{\mathrel{\raise.3ex\hbox{$>$\kern-.75em\lower1ex\hbox{$\sim$}}}}
\def\proptwid{\mathrel{\raise.3ex\hbox{$\propto$\kern-.75em\lower1ex\hbox{$\sim$}}}}

\newcommand{\IAIFI}{The NSF AI Institute for Artificial Intelligence and Fundamental Interactions}
\newcommand{\MIT}{Department of Physics and Kavli Institute for Astrophysics and Space Research, Massachusetts Institute of Technology, 77 Massachusetts Avenue, Cambridge, MA 02139, USA}
\newcommand{\CfA}{Center for Astrophysics $|$ Harvard \& Smithsonian, Cambridge, MA 02138, USA}

\defcitealias{TMS}{TMS}
\setcitestyle{notesep={; }}

\begin{document}

\title{\textbf{When IIb Ceases To Be: Bridging the Gap Between IIb and Short-plateau Supernovae }} 
\shorttitle{SN 2023wdd and SN 2022acrv}

\author[0000-0003-4914-5625]{Joseph Farah}
\affiliation{Las Cumbres Observatory, 6740 Cortona Drive, Suite 102, Goleta, 
CA 93117-5575, USA}
\affiliation{Department of Physics, University of California, Santa Barbara, 
CA 93106-9530, USA}
\author[0000-0003-4253-656X]{D. Andrew Howell}
\affiliation{Las Cumbres Observatory, 6740 Cortona Drive, Suite 102, Goleta, 
CA 93117-5575, USA}
\affiliation{Department of Physics, University of California, Santa Barbara, 
CA 93106-9530, USA}
\author[0000-0002-1125-9187]{Daichi Hiramatsu}
\affiliation{Center for Astrophysics \textbar{} Harvard \& Smithsonian, 60 Garden Street, Cambridge, MA 02138-1516, USA} \affiliation{The NSF AI Institute for Artificial Intelligence and Fundamental Interactions, USA}
\author[0000-0001-5807-7893]{Curtis McCully}
\affiliation{Las Cumbres Observatory, 6740 Cortona Drive, Suite 102, Goleta, 
CA 93117-5575, USA}
\author[0000-0002-1895-6639]{Moira Andrews}
\affiliation{Las Cumbres Observatory, 6740 Cortona Drive, Suite 102, Goleta, 
CA 93117-5575, USA}
\affiliation{Department of Physics, University of California, Santa Barbara, 
CA 93106-9530, USA}
\author[0000-0001-9570-0584]{Megan Newsome}
\affiliation{Las Cumbres Observatory, 6740 Cortona Drive, Suite 102, Goleta, 
CA 93117-5575, USA}
\affiliation{Department of Physics, University of California, Santa Barbara, 
CA 93106-9530, USA}
\author[0000-0003-0209-9246]{Estefania Padilla Gonzalez}
\affiliation{Las Cumbres Observatory, 6740 Cortona Drive, Suite 102, Goleta, 
CA 93117-5575, USA}
\affiliation{Department of Physics, University of California, Santa Barbara, 
CA 93106-9530, USA}
\author[0000-0002-7472-1279]{Craig Pellegrino}
\affiliation{Department of Astronomy, University of Virginia, Charlottesville, VA 22904, USA}
\author[0000-0002-9392-9681]{Edo Berger}
\affiliation{\CfA}
\affiliation{\IAIFI}
\author[0000-0003-0526-2248]{Peter Blanchard}
\affiliation{\CfA}
\author[0000-0001-6395-6702]{Sebastian Gomez}
\affiliation{\CfA}
\author[0000-0003-0871-4641]{Harsh Kumar}
\affiliation{\CfA}
\author[0000-0002-4924-444X]{K. Azalee Bostroem}
\affiliation{Steward Observatory, University of Arizona, 933 North Cherry Avenue, Tucson, AZ 85721-0065, USA}
\author[0000-0003-3656-5268]{Yuan Qi Ni}
\affiliation{Las Cumbres Observatory, 6740 Cortona Drive, Suite 102, Goleta, 
CA 93117-5575, USA}
\affiliation{Department of Physics, University of California, Santa Barbara, 
CA 93106-9530, USA}

\affiliation{Kavli Institute for Theoretical Physics, University of California, Santa Barbara, CA 93106, USA}
\author[0000-0003-4906-8447]{A.~Gagliano}
\affiliation{\IAIFI}
\affiliation{\CfA}
\affiliation{\MIT}
\author[0000-0002-7352-7845]{Aravind P. Ravi}
\affiliation{Department of Physics and Astronomy, University of California, Davis, 1 Shields Avenue, Davis, CA 95616-5270, USA}

\shortauthors{Farah et al.}

\correspondingauthor{$^\dag$Joseph R. Farah}
\email{josephfarah@ucsb.edu}

\begin{abstract}
Hydrogen-rich supernovae (SNe) span a range of hydrogen envelope masses at core collapse, producing diverse light curves from extended plateaus in Type IIP SNe to double-peaked Type IIb SNe. Recent hydrodynamic modeling predicts a continuous sequence of light-curve morphologies as hydrogen is removed, with short plateau SNe (plateau durations $\approx50\text{--}70$ days) emerging as a transitional class. However, the observational boundary between IIb and short-plateau remains poorly defined, and thus far unobserved. We report on extensive photometric and spectroscopic follow-up of SN 2023wdd and SN 2022acrv, candidate transitional events on the low-mass end of the short-plateau class. Both exhibit weak, double-peaked light curves which we interpret as exceptionally short plateaus (10–20 days), and hybrid spectral features: persistent H$\alpha$ absorption with He I contamination, but without the helium dominance characteristic of IIb SNe. Using analytic shock-cooling models and numerical light curve fitting, we estimate hydrogen-rich envelope masses of ${\sim}0.6\text{--}0.8\ M_\odot$--significantly larger than canonical IIb values ($\lesssim0.1 \ M_\odot$) but consistent with the ${\sim}0.9\ M_\odot$ threshold predicted for short-plateau behavior. Although the progenitor radii inferred from analytic and numerical methods differ by factors of 2–5, envelope mass estimates are consistent across approaches.
Comparisons to well-studied IIb (SN 2016gkg, SN 2022hnt), short-plateau (SN 2023ufx, SN 2006ai, SN 2016egz, SN 2006Y), and IIP SNe (SN 2023ixf, SN 2013ej) suggest a monotonic relationship between hydrogen envelope mass and plateau length consistent with analytic and numerical expectations. These findings provide additional evidence for a continuous distribution of envelope stripping in hydrogen-rich core-collapse progenitors and place SN 2023wdd and SN 2022acrv along the IIb/short-plateau boundary.
\end{abstract}


\section{Introduction}
\label{sec:introduction}

Hydrogen-rich supernovae (SNe) represent some of the most diverse outcomes of massive ($>8 \ M_\odot$) star evolution, characterized by the removal—partial to near-complete—of the hydrogen and/or helium shells prior to core collapse. Within this broad umbrella, Type IIb SNe retain a thin layer of hydrogen, manifesting as early-time hydrogen lines in their spectra that later give way to helium-dominated features \citep[e.g.,][]{Woosley1994,Filippenko1997,Modjaz2019}. At the opposite extreme lie the Type IIP SNe, which hold onto a substantial hydrogen envelope and exhibit long ($\sim$100 days) photometric plateaus \citep[see][for a review]{Branch2017}. Historically, this distinction between unstripped IIP and mostly-stripped IIb events suggested that these were separate pathways, reflecting either minimal or extensive mass loss. Yet, an emerging body of theoretical and observational evidence indicates the possibility of a continuum in the amount of hydrogen envelope stripping, raising the question of how many physically intermediate (or ``transitional'') events remain unclassified under the standard taxonomy \citep[e.g.,][]{Chornock2011,Gal-Yam2017,Hiramatsu2021}.

The conventional picture of Type IIb progenitors invokes binary interaction or strong stellar winds to peel away nearly all the hydrogen above the helium core, leaving behind an envelope of only $\sim0.01-0.1 \ M_\odot$ of H-rich material \citep{Smith2014,Sravan2019,Bersten16gkg}. Double-peaked IIb supernovae often display prominent early-time excess emission in their light curves, where an initial shock-cooling peak gives way to a re-brightening powered by radioactive nickel-56 decay \citep[e.g.,][]{Richmond1994,Arcavi2011,Kumar2013,Bufano2014,Morales-Garoffolo2014,Fremling2019,Armstrong2021,Pellegrino2023,Farah2025}. By contrast, Type IIP supernovae—commonly associated with red supergiant progenitors—experience only modest envelope stripping and thus show extended plateaus for 80--120 days, followed by a transition to the radioactive tail \citep[e.g.,][]{Valenti2014,Valenti2016,Hosseinzadeh2023}. Between these two archetypes lie the so-called short-plateau SNe, showing plateaus of only $\sim50\textrm{--}70$ days, or sometimes even shorter \citep{Hiramatsu2021}. These objects appear to occupy an intermediate regime in terms of hydrogen envelope mass $\sim1.0-2.9 \ M_\odot$, suggesting the boundaries between ``IIb'' and ``IIP'' do not form sharp categories, but rather reflect a distribution of envelope retention.

A central open question, therefore, is whether the observed range of hydrogen envelope masses--spanning canonical Type IIP (several $M_\odot$) down to the thin envelopes of IIb supernovae--arises along a smooth continuum of stripping.  Such a continuum view would help clarify the role of binary interactions, mass-transfer episodes, stellar winds, or eruptive mass loss in shaping the final envelope states of massive stars \citep{Claeys2011,Smith2014,Ouchi2017,Schneider2021,Ercolino2024}. Systematic surveys targeting the earliest phases of core-collapse events are increasingly revealing transients with partial hydrogen depletion \citep[e.g.,][etc.]{Anderson2014,Dong2024,ayala2025earlylightcurveexcess}. Yet the supporting data are still sparse, especially around the boundary between Type IIb and the short-plateau regime.

Motivated by these considerations, we present detailed observations of two supernovae--SN 2023wdd and SN 2022acrv--that exhibit photometric and spectroscopic signatures intermediate to double-peaked IIb and short-plateau events. Their double-peaked light curves bear a qualitative resemblance to IIb SNe, but the initial drop is far more subtle, and a broad H$\alpha$ persists longer than archival IIb SNe. In \autoref{sec:boundary_objects}, we review past examples of ``boundary'' objects across various core-collapse supernova types. In \autoref{sec:data}, we report on the follow-up campaign and spectrophotometric reductions for both SN 2023wdd and SN 2022acrv. In \autoref{sec:spectroscopic_analysis} we review the spectroscopic evolution of both objects. In \autoref{sec:modeling}, we present the photometric evolution and model the light curve to measure the hydrogen-rich envelope masses of both transients, contextualizing them along the continuum of hydrogen-rich objects. In \autoref{sec:comparisons}, we draw comparisons in evolution and inferred parameters to other supernovae in the continuum. Finally, in \autoref{sec:conclusions}, we summarize our results. 

\section{On Boundary objects}
\label{sec:boundary_objects}

Historically, classifications of SNe have been made based on the presence of specific spectral lines, promoting placement of SNe into mutually exclusive categories which correspond to particular progenitor or explosion channels \citep{Filippenko1997,Gal-Yam2017}. Recently, denser and deeper transient surveys have helped identify relatively rare SNe which display features characteristic of multiple categories, challenging the exclusivity of previously established categories and suggesting the presence of a continuum of features and physical processes driving them \citep{Bellm2019,Tonry2018,Brown2013}. These objects may lie along the boundary of two categories in one more properties (``boundary objects''), making straightforward classification difficult. 

Among core-collapse SNe--which have been historically categorized (i.e., IIP, IIL, IIb, Ib, Ic) by the amount of hydrogen in their spectra--a number of subclassifications have arisen, partially due to the discovery of boundary objects. The boundary between Type II SNe (core-collapse with hydrogen) and Type Ib SNe (core-collapse without hydrogen) was challenged by the discovery of SN 1993J, which had a prominent H$\alpha$ feature associated with Type II SNe that later disappeared, leaving a clear Type Ib spectrum \citep{2000AJ....120.1487M,Woosley1994}. This motivated the creation of a new classification of SNe--Type IIb SNe--which have most (but not all) of their hydrogen stripped, leading to a hydrogen-rich-envelope mass $M_{\textrm{H}_{\textrm{env}}}\ll 1.0 M_\odot$ at the time of explosion, in sharp contrast to Type IIP SNe ($M_{\textrm{H}_{\textrm{env}}}\gg 1.0 M_\odot$) \citep{Filippenko1997}. This new categorization was recently challenged yet again, with the discovery of SNe with shorter plateaus than previously studied Type IIP SNe apparently caused by a hydrogen-rich envelope mass of $M_{\textrm{H}_{\textrm{env}}}\sim 2.0 M_\odot$, $\sim5-10$x larger than other IIb SNe but $2-5$x smaller than ordinary Type IIP SNe \citep{Hiramatsu2021}. \cite{Hiramatsu2021} proposes a new category--the short-plateau SNe--to categorize supernovae with $0.91 \ M_\odot \lesssim M_{\textrm{H}_{\textrm{env}}} \lesssim 2.1 \ M_\odot$, resulting in a plateau $\lesssim80-100$ days in length, and demonstrates that three supernovae (SN 2006ai, SN 2016egz, SN 2006Y) seem to be well-classified by this new category, lying along the boundary between Type IIP SNe and short-plateau SNe.

In this work, we seek to demonstrate that the lower boundary of the short-plateau SNe subclass--between IIb and short-plateau SNe--may be part of a continuum of hydrogen stripping, and not a rigidly defined boundary. Observations of legacy supernovae have suggested that the hydrogen stripping present in Type IIb SNe may be part of a spectrum connecting to Type Ib SNe  \citep{Dong2024,Teffs2020}. Numerical models have extended this spectrum from Type IIb SNe to Type IIL and Type IIP SNe, suggesitng that Type II SNe have experienced hydrogen loss to varying degrees, providing further evidence of a continuum \citep{Hiramatsu2021,Fang2025}. Below, we present observations and analysis of two new supernovae (SN 2023wdd and SN 2022acrv) which appear to lie along the IIb/short-plateau boundary, with $M_{\textrm{H}_{\textrm{env}}}\sim 1.0 M_\odot$, strengthening the continuum argument for hydrogen stripping. 
\section{Observations and data}
\label{sec:data}

\subsection{SN 2023wdd}
\label{sub:sn_2023wdd}

SN 2023wdd was discovered at $o$-band magnitude 17.9 by the Asteroid Terrestrial Last-impact Alert System \citep[ATLAS]{2023TNSTR2790....1T} on Nov 1, 2023 (MJD 60249). This discovery was automatically reported to the Transient Name Server (TNS) on the same day as discovery. SN 2023wdd was located at right ascension (R. A.) 01:18:11.045 and declination (decl) +38:26:32.75. The SN is associated with host spiral galaxy UGC 00831. SN 2023wdd was initially classified as a young Type II SN by Li-Jiang One hour per Night observation of SNe (LiONS) \citep{Li2023}, before being reclassified by the extended Public ESO Spectroscopic Survey of Transient Objects (ePESSTO+) on Nov 6 as a Type IIb \citep{2023TNSCR2869....1M}. A broad H$\alpha$ emission feature constrains the redshift to $z\sim 0.024$, consistent with the host galaxy redshift of $z\sim0.024303$ (value from NED\footnote{The NASA/IPAC Extragalactic Database (NED) is funded by the National Aeronautics and Space Administration and operated by the California Institute of Technology. }). Based on a redshift of $z\sim0.024$ and assuming a flat universe ($\Omega_M = 0.286, \Omega_{\Lambda}=0.714$) with $H_0\approx69.6 \mathrm{km \ s^{-1} \ Mpc^{-1}}$, we adopt a distance of 105.3 Mpc to the SN. Based on the \cite{Schlafly2011} dust map, we adopt a line-of-sight extinction value of $E(B-V)_{\rm MW} = 0.0373 \pm 6\times10^{-4}$ mag. We correct all light curve and photometric data presented in this paper with this extinction value. Next, we describe our multiband follow-up observations of SN 2023wdd. 

Shortly ($\lesssim 2$ days) after discovery, we triggered a high-cadence, multiband follow-up campaign with the Las Cumbres Observatory \citep[LCO,][]{Brown2013}. As part of the Global SN Project (GSP), LCO provides a global, networked array of robotic telescopes designed specifically for rapid follow-up of active transients. We obtained photometric observations in $U, B, V, g, r$ and $i$ bands using the LCO 1m and 2m telescopes. All photometric LCO data were reduced using the \texttt{lcogtsnpipe} photometric reduction pipeline, which uses point-spread-function (PSF) and color term computation before magnitude extraction and measurement \citep{Valenti2016}. Extracted magnitudes were then calibrated to Landolt standards (using Vega magnitudes) for $UBV$ filters \citep{Landolt2009} and the Sloan Digital Sky Survey (SDSS) catalog (using AB magnitudes) for the remaining bands \citep{Smith2002}. 

The LCO observatory network was also used to obtain high-cadence spectra for the transient. We obtained seventeen spectra between day 2 and day 88 using the FLOYDS spectrograph, which is mounted on the northern 2m LCO site (Haleakala Observatory). This spectrograph has a wide coverage (350 nm to 1 micron), and data is processed via the custom LCO \texttt{floydsspec} pipeline \citep{Valenti2016}. \texttt{floydsspec} incorporates cosmic ray flagging, spectrum extraction, and calibration. LCO spectra were used to identify host galaxy extinction via the method of \cite{Poznanski2012}, which fits the equivalent width of the Na I ($\lambda5893$) doublet to estimate extinction. We find a best-fit extinction coefficient of $E(B-V)_{\rm host}=0.105 \pm 0.022$ mag. We additionally correct all light curve and photometric data presented in this paper with this host extinction value.

In addition to the LCO follow-up campaign, we also included forced photometry from the ATLAS survey \citep{Tonry2018}. ATLAS is a pair of two 0.5m all-sky survey telescopes which observe in at least one band at least once every other day. The ATLAS instruments are located in Hawaii, on Mauna Loa and Haleakala. ATLAS observed SN 2023wdd on $\sim$dozen epochs in multiple bands simultaneously during the LCO observation campaign. ATLAS photometry was obtained in both the custom ATLAS $o$- (orange) and $c$- (cyan) band filters. We reduced all ATLAS photometry using the ATLAS forced photometry pipeline\footnote{\href{https://fallingstar-data.com/forcedphot/}{https://fallingstar-data.com/forcedphot/}}, with calibration to the Pan-STARRS catalog. 

Besides LCO and ATLAS, we also incorporated publicly-available observations made by the Zwicky Transient Facility \citep[ZTF]{Bellm2019}. ZTF is a nightly transient survey camera incorporated into the 48-inch Palomar Observatory telescope. We incorporate observations of SN 2023wdd (ZTF23abobwsd) made between Oct 29, 2023 and Jan 06, 2024. Subtractions and magnitude extractions are automatically performed using templates previously collected by ZTF and processed by the ZTF photometry reduction pipeline \citep{Masci2023}. 

\begin{figure}
    \centering
    \includegraphics[scale=0.6]{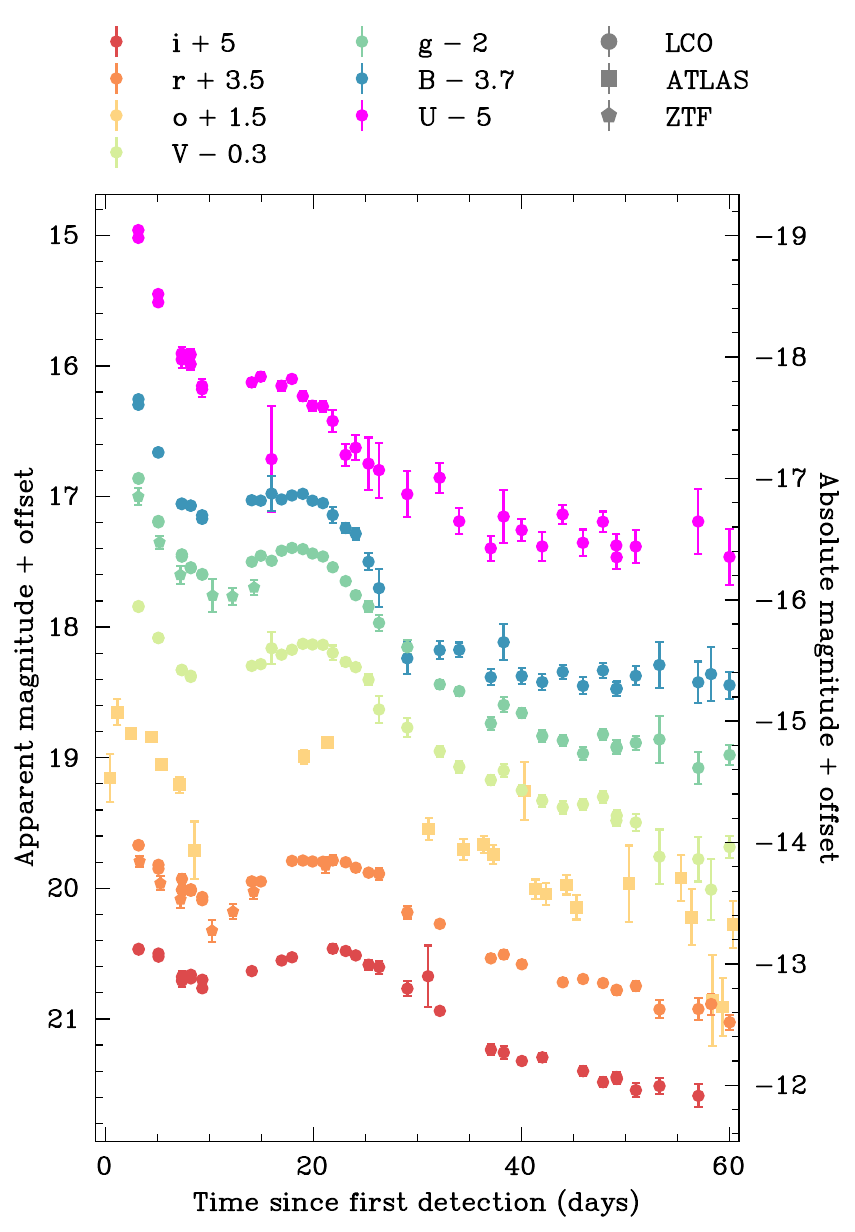}
    \caption{Full, multiband photometric data collected on SN 2023wdd by LCO, ATLAS, and ZTF. Follow-up programs for all three facilities clearly captured the initial decline and rise to second peak. We offset both absolute and apparent magnitudes by arbitrary amounts (shown in legend) to improve visibility. Error bars indicate $1\sigma$ uncertainty. Vertical gray lines indicate when spectra were taken. The estimated explosion epoch is based on the shock cooling fits discussed in \autoref{sec:modeling}.}
    \label{fig:23wdd_full_lc}
\end{figure}

\begin{figure}
    \centering
    \includegraphics[scale=0.6]{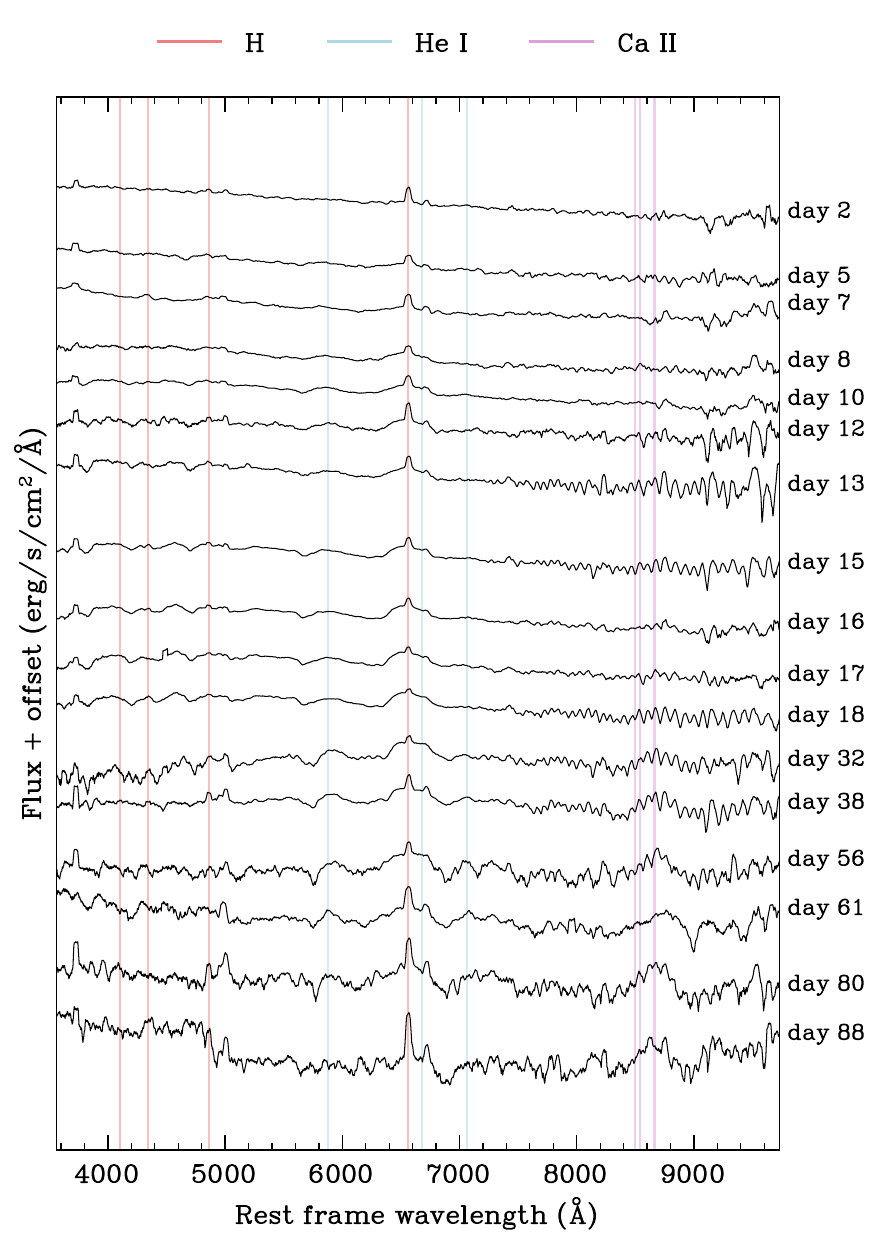}
    \caption{The spectral evolution of SN 2023wdd from explosion through the first 88 days. All spectra shown are from the FLOYDS instrunment at LCO. We highlight the rest frame wavelengths of hydrogen (red line), helium (blue line) and calcium II (purple line). The persistent narrow feature at the hydrogen rest wavelength is contamination from the host galaxy (as evidenced by its unchanging equivalent width, see \autoref{sub:csm_investigation} for a discussion), superimposed over the broad SN hydrogen emission line. At later times, there is weak contamination of the broad hydrogen from a coincident helium absorption, but the hydrogen feature is still significant 88 days post-explosion.}
    \label{fig:23wdd_spectra}
\end{figure}

A compilation of all photometry data collected for SN 2023wdd is shown in \autoref{fig:23wdd_full_lc}. A compilation of all spectra for SN 2023wdd is shown in \autoref{fig:23wdd_spectra}.

\subsection{SN 2022acrv}
\label{sub:sn_2022acrv}

SN 2022acrv was discovered at $o$-band magnitude 18.3 by ATLAS, and reported to the Transient Name Server (TNS) on Dec 12, 2022. SN 2022acrv was located at R.A. 05:46:44.992 and decl -20:08:58.66. The SN was not initially associated with any visible host, but may be associated with GALEXASC J054644.80-200858.1. SN 2022acrv was classified as a young Type II SN by ePESSTO+ within two weeks of discovery \citep{Gkini2022}. A H$\alpha$ emisison feature constrains the redshift to $z\sim 0.028$, consistent with the candidate host galaxy redshift of $z\sim0.029067$ (value from NED). Based on a redshift of $z\sim0.0291$ and the cosmological parameters used above, we adopt a distance of 127.7 Mpc to the SN. Based on the \cite{Schlafly2011} dust map, we adopt a line-of-sight extinction value of $E(B-V)_{\rm MW} = 0.0441 \pm 3\times10^{-4}$ mag. We correct all light curve and photometric data presented in this paper with this extinction value. Next, we describe our multiband follow-up observations of SN 2022acrv. 

\begin{figure}
    \centering
    \includegraphics[scale=0.6]{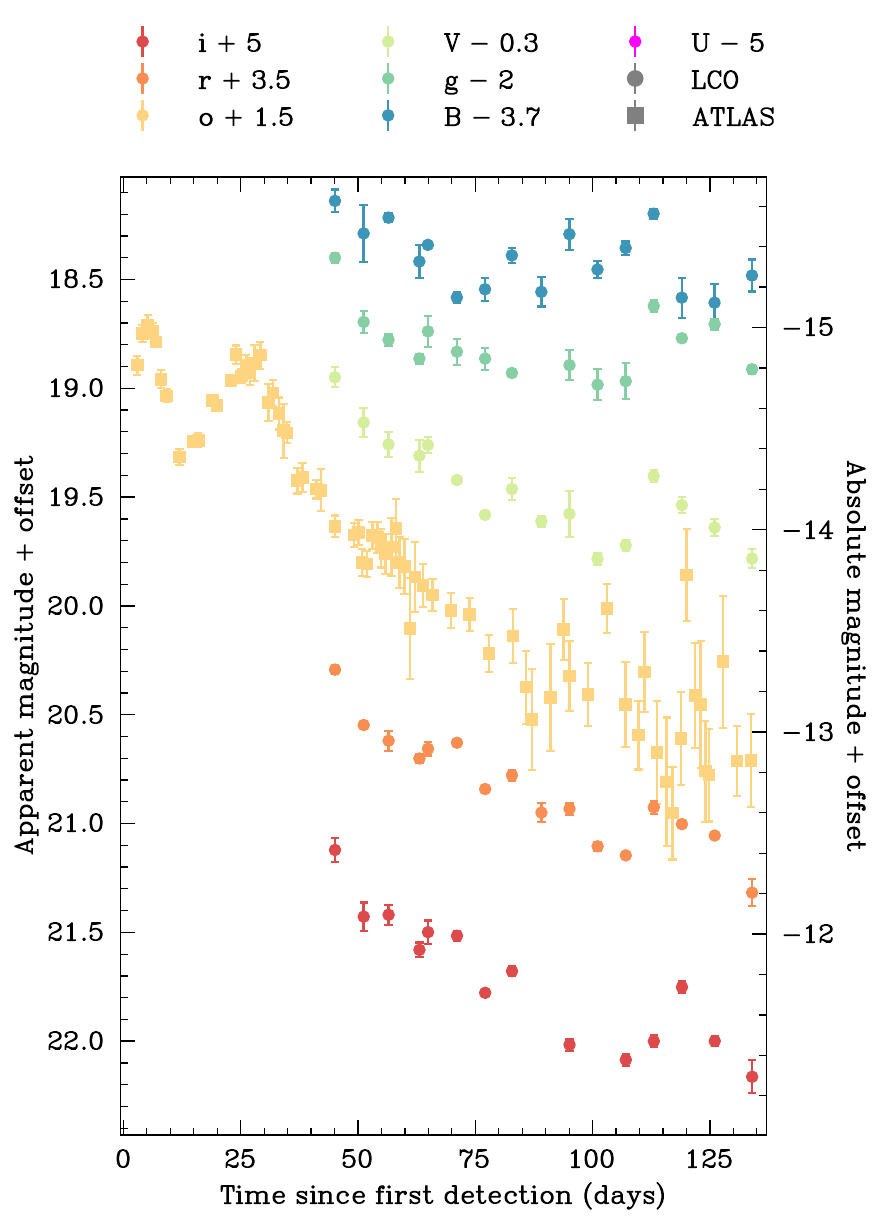}
    \caption{Full, multiband photometric data collected on SN 2022acrv by LCO and ATLAS. We offset both absolute and apparent magnitudes by arbitrary amounts (shown in legend) to improve visibility.  Error bars indicate $1\sigma$ uncertainty. Vertical gray lines indicate when spectra were taken. The estimated explosion epoch is based on the shock cooling fits discussed in \autoref{sec:modeling}.}
    \label{fig:22acrv_full_lc}
\end{figure}

\begin{figure}
    \centering
    \includegraphics[scale=0.65]{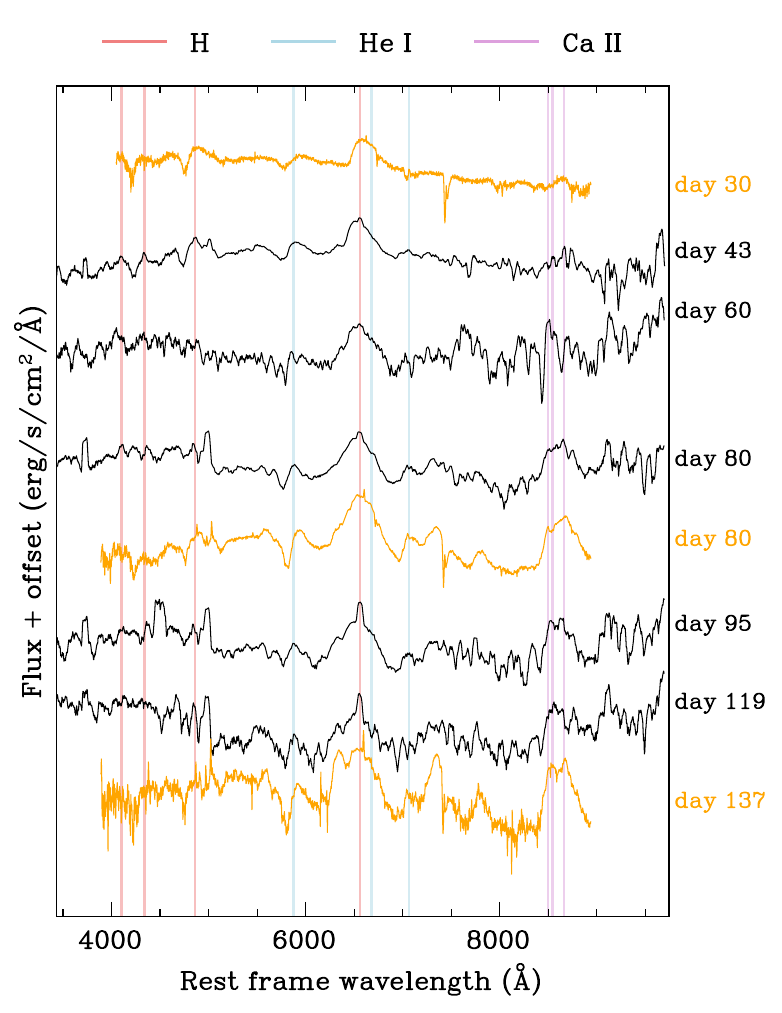}
    \caption{The spectral evolution during the first 137 days of the evolution of SN 2022acrv. Spectra from LCO are shown in black, while spectra from the Magellan instrument are shown in orange. We highlight the rest frame wavelengths of hydrogen (red line), helium (blue line) and calcium II (purple line). The persistent narrow feature at the hydrogen rest wavelength is contamination from the host galaxy, superimposed over the broad SN hydrogen emission line.}
    \label{fig:22acrv_spectra}
\end{figure}

LCO was not triggered on this object until well into its evolution ($\approx$ 1 month from discovery). By this point, the transient had executed a initial rise, partial decline, and secondary rise to peak, as visualized in \autoref{fig:22acrv_full_lc}. LCO obtained weekly photometry and spectra on SN 2022acrv from Jan 21, 2023 until the object faded $\approx$100 days later, providing a detailed multi-band light curve of the SN tail. LCO photometric and spectroscopic reduction details are identical to \autoref{sub:sn_2023wdd}, with one exception: as the object was located in the southern sky, we employed the FLOYDS spectrograph mounted on the 2m telescope at Siding Spring Observatory in Australia. In addition to FLOYDS spectra, we also obtained spectra using the Low Dispersion Survey Spectrograph 3 \citep[LDSS-3][]{Stevenson2016} on the Magellan Clay Telescope at the Las Campanas Observatory in Chile, which covers primarily optical wavelength range (3700 -- 10060 angstroms). Spectra were processed following the procedure described in \cite{Hiramatsu2024}: extraction and reduction were performed via PyRAF, with calibration to a standard star spectrum taken within one week of each target spectrum. The Magellan and FLOYDS spectra of SN 2022acrv are shown in \autoref{fig:22acrv_spectra}.

As with SN 2023wdd, we used the available Magellan spectra to identify host galaxy extinction via the method of \cite{Poznanski2012}. We find a best-fit extinction coefficient of $E(B-V)_{\rm host}=0.134\pm0.037$ mag. We additionally correct all light curve and photometric data presented in this paper with this host extinction value.

Finally, similarly to SN 2023wdd, we incorporate every-other-day ATLAS observations in $o$- and $c-$ band, available via the forced photometry server. See \autoref{sub:sn_2023wdd} for reduction details.

\section{Spectroscopic evolution}
\label{sec:spectroscopic_analysis}

Both SN 2023wdd and SN 2022acrv show similar spectral evolution, with particular features supporting their characterization as candidates for boundary objects between Type IIb and short-plateau SNe. 

The full spectral evolution of SN 2023wdd is shown in \autoref{fig:23wdd_spectra}. LCO spectra are available between day 2 and day 88 (days from explosion epoch measured using the shock cooling fits in \autoref{sec:modeling}). The earliest spectra (within 1 week of explosion) show an almost featureless blue continuum, with the exception of a persistent narrow hydrogen feature produced by the host galaxy (see \autoref{sub:csm_investigation} for further discussion of the origin of this line). The lack of features is due to the extreme temperature of the ejecta \citep{Filippenko1997}, which we estimate from blackbody fits to photometry to be $\gtrsim10^4$ K in the first few days of the observations. As the ejecta cool, lines begin to show, especially a broad H$\alpha$ feature at $\approx6500$ angstroms, facilitating classification of this object as a Type II SN. The broad H$\alpha$ line does not appear to have an associate P Cygni absorption feature, which has been discussed in previous literature as evidence of interaction with circumstellar material (CSM) \citep{Filippenko1997,Hillier2019,Dessart2022}--we consider this possibility in \autoref{sub:csm_investigation}. The hydrogen line then develops a flat-top due to contamination from a coincident helium line, which, in conjunction with the double-peaked light curve, suggests a IIb SN classication \citep{Filippenko1997,Gal-Yam2017}. 

\begin{figure*}
    \centering
    \hspace*{-0.5cm}
    \includegraphics[scale=0.75]{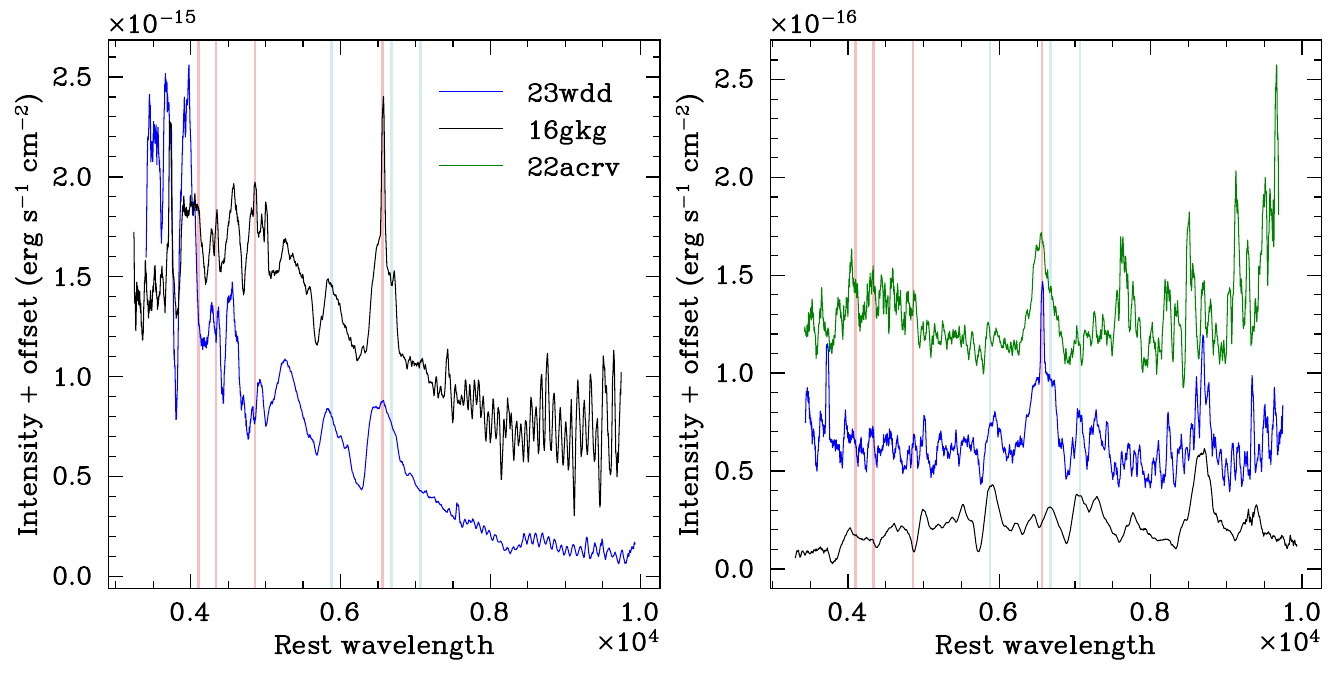}
    \caption{Comparison of the spectra of our boundary objects versus a well-studied Type IIb SNe, SN 2016gkg (black). Specific elements are shown using the same legend as \autoref{fig:23wdd_spectra} and \autoref{fig:22acrv_spectra}. \textit{(left)} Comparison of the spectra of SN 2023wdd and SN 2016gkg $\sim2$ weeks into their evolution (SN 2022acrv does not have spectra available this early). At this early time, the hydrogen line is still present, and developing the flat-top associated with helium contamination \citep{Filippenko1997}. Even at this early stage, the helium contamination in SN 2016gkg is more prominent than in SN 2023wdd. \textit{(right)} Comparison of the spectra of SN 2016gkg, SN 2023wdd and SN 2022acrv $\approx 50$ days into their respective evolutions. At this stage, the heavily-stripped envelope of SN 2016gkg has recombined its limited hydrogen, causing the previously prominent H$\alpha$ feature to be overwhelmed by the strong coincident helium line that was causing the flat-top in earlier spectra. By contrast, SN 2023wdd displays some helium contamination in the form of a flat-top, but both SN 2023wdd and SN 2022acrv still have significant H$\alpha$ features, suggesting that their envelopes contain substantially more hydrogen than SN 2016gkg ($M_{\textrm{H}_{\textrm{env}}}\sim 0.01 \ M_\odot$).}
    \label{fig:spec_comp}
\end{figure*}

However, in a typical double-peaked IIb SN, the H$\alpha$ line is usually weak, and quickly gets contamined by the growing helium absorption feature, made visible by stripping of the hydrogen-rich outer envelope of the progenitor star \citep{Filippenko1997,2000AJ....120.1487M,Pellegrino2023,Farah2025}. In contrast to this expectation, the broad H$\alpha$ remains fairly strong even almost 100 days post-explosion; and even though the presence of a He I line coincident with H$\alpha$ contaminates the H$\alpha$ emission somewhat, it does not dominate it as it might in a normal IIb SN. The He I feature--visible as a P cygni at $\approx 5800$ angstroms--is also weaker than expected for a IIb. To emphasize the distinction with an ordinary IIb SNe, we compare SN 2023wdd at early and late times to a well-known Type IIb SN, SN 2016gkg, in \autoref{fig:spec_comp}. At early times, both spectra show a prominent H$\alpha$ feature with moderate helium contamination, but by $\sim2$ months in, SN 2023wdd has retained its H$\alpha$ feature, while the H$\alpha$ feature in SN 2016gkg almost entirely disappears, making the spectrum look more similar to a Type Ib SN.

The evolution of SN 2022acrv has some overlap in these peculiarities, and is shown in full in \autoref{fig:22acrv_spectra}. Our hybrid (LCO + Magellan) spectroscopic coverage is available from day 30 to day 137. Despite the clear double-peak in the light curve of SN 2022acrv (suggesting a IIb SN), the spectra is dominated for at least the first 137 days by a broad H$\alpha$. Indeed, despite the presence of helium in the ejecta (visible at all epochs at $\approx 5800$ angstroms), the central H$\alpha$ is not significantly contaminated, conflicting with the classification as a IIb SN. Like SN 2023wdd, this peculiar mismatch between the light curve and spectra motivates investigation of SN 2022acrv as a boundary object between Type IIb and short-plateau SNe. 

We also obtained a late (364 days post-explosion) spectrum of the SN 2022acrv host galaxy using Magellan, in order to assess metallicity of host galaxy and therefore the immediate environment of the SN. Following \cite{Hiramatsu2021}, we recover line fluxes by fitting a Gaussian with a vertical offset to the following lines: H$\alpha$ ($\lambda$6563), H$\beta$ (4861), O II ($\lambda$3727), O III ($\lambda$4959),  N II ($\lambda$6583), S II ($\lambda$6717), S II ($\lambda$6731). Using the derived expresssions of \cite{Pettini2004} (specifically equations 2 and 3), we estimate an O3N2 value of $\approx 1.938$ and an N2 value of $\approx -1.457$, which corresponds to a host galaxy metallicity of $Z \approx 0.255 \ Z_\odot \pm 0.01 Z_\odot$. We obtain a consistent host galaxy metallicity using the open-source tool \texttt{PyMCZ} \citep{PyMCZ}. This metallicity is consistent with the small sample of other short-plateau SNe, and is most similar to SN 2016egz \citep[$Z\sim0.27 \ Z_\odot$][]{Hiramatsu2021}. Along with the host galaxy of SN 2023ufx, which had an extremely low metallicity ($Z \lesssim 0.05 \ Z_\odot$), short-plateau SNe seem to moderately favor lower-metallicity galaxies, though the sample of short-plateau SNe is too small to make this assertion with any confidence. The host metallicities of these boundary objects ($12+\log(O/H)\approx8.2$) are also consistent with previous studies of IIb SNe host galaxy metallicities \citep{Galbany2016}, which found that IIb SNe are preferentially found in local environments with lower metallicities ($12+\log(O/H)\lesssim8.5$).

\section{Light curve modeling}
\label{sec:modeling}

\subsection{Photometric evolution}
\label{sub:photometric_evolution}

The full photometric evolution of SN 2023wdd is available in \autoref{fig:23wdd_full_lc}. The combined LCO, ATLAS, and ZTF light curves provide a densely sampled, multiband reconstruction of the transient evolution, from a single-color constraint on the rise to the radioactive tail. The luminosity and temperature evolution along the first $\approx10$ days is consistent with the nominal predictions of a shocked envelope expanding and cooling \citep[$T\propto t^{-0.45}$,][]{Rabinak2011}. Around $\approx 10\text{--}20$ days, the light curve executes a weak double-peak, with $\lesssim0.5$ mag difference in brightness from dip to second peak in all bands. After rising to a second peak at $\approx22$ days, the light curve relaxes onto an apparent radioactive tail. Modeling the luminosity along this tail by the radioactive decay of nickel-56 \citep{Arnett1980,Lyman2016} yields an estimated nickel mass synthesized during explosion of $\approx0.039 \ M_\odot$. 

The full photometric evolution of SN 2022acrv is similar to SN 2023wdd, and is shown in \autoref{fig:22acrv_full_lc}. A $\approx$ daily reconstruction of the transient behavior is only available in ATLAS $o$- and $c$-band, retrieved via forced photometry. In the ATLAS data, two peaks are clearly observed, with multiple data points constraining the initial rise. Due to the far more limited multicolor data at early times, only a few epochs of bolometric information are available. Nonetheless, the first peak of SN 2022acrv is consistent with the predictions of shock cooling up to $\approx9$ days \citep{Rabinak2011}. After executing the shock cooling peak and decline, the transient rises to a secondary nickel peak before relaxing onto a radioactive tail, which is well-probed in multiple filters via a combination of LCO and ATLAS photometry.  Modeling the luminosity along this tail by the radioactive decay of nickel-56 using the same approach as above yields an estimated nickel mass synthesized during explosion of $\approx0.064 \ M_\odot$. 

The weak double-peak light curve feature indicates both objects have experienced some stripping of their outer hydrogen-rich envelopes, supported by the helium contamination of the H$\alpha$ line discussed in \autoref{sec:spectroscopic_analysis}. In the sections below, we attempt to use the light curve evolution described above to constrain properties of the progenitor system and explosion for each object. In particular, we seek a constraint of the hydrogen envelope mass to further solidify the placement of each supernova along the continuum of Type II/stripped-envelope objects.

\subsection{CSM investigation}
\label{sub:csm_investigation}

Interaction between expanding ejecta and a high-density CSM can power light curve luminosities comparable to shock cooling and radioactive nickel decay, creating degeneracies in the light curve that can complicate parameter inference. We examine possible indications of CSM in SN 2023wdd and SN 2022acrv observations to assess whether CSM interaction is significant enough to warrant dedicated modeling in our parameter inference scheme. SN 2022acrv does not display signs of significant CSM interaction in its light curve or spectra; by contrast, SN 2023wdd has two potential indicators. These are: (i) an apparent lack of a P Cygni profile in the H$\alpha$ line, which has been linked to interaction in previous studies \citep{Filippenko1997,Hillier2019,Dessart2022}, and (ii) an extremely narrow and strong H$\alpha$ component, which may be an indication of CSM interaction or, alternatively, may be host-galaxy contamination. We consider both of these indicators below. 

First, we consider the apparent lack of a P Cygni profile in the H$\alpha$ line. To investigate, we model the H$\alpha$ as the combination of a linear underlying continuum, a P Cygni profile and a narrow Gaussian emission line profile. We construct the P Cygni profile as separate absorption and emission components, following e.g., \cite{Foley2013} and \cite{Kwok2025}. We find that once that the ejecta have cooled sufficiently (1-2 weeks), the H$\alpha$ line indeed develops a clear P Cygni component, with an absorption feature $\sim1/2$ the strength of the coupled emission feature. Using the method of \cite{Castor1979} and \cite{Lamers1987}, we set an upper limit of $\rho_{\textrm{CSM}}\sim 8\times10^{-16} \mathrm{g \ cm^{-3}}$ ($\dot{M} \sim 1.5\times10^{-4} \mathrm{M_\odot \ yr^{-1}}$). This density is not sufficient to warrant additional modeling, as, based on the results of \cite{Hiramatsu2021}, such a density is not large enough to impact the light curve evolution significantly. 

Next, we examine the strong, narrow feature coincident with the H$\alpha$ line. This line is likely due to host contamination, given the presence of other prominent host galaxy lines (e.g., H$\beta$, [O III], [S II]). If this narrow feature is produced via interaction with a CSM, it may vary as the ejecta expands, like in a Type IIn SN; by contrast, if the feature is produced due to host contamination, it should be unchanged as the supernova evolves. We measure the equivalent width of the narrow H$\alpha$ feature over time and do not observe statistically significant variation over the course of the observations. Fitting a line to the variation yields a slope statistically consistent with zero. Therefore, we attribute the narrow H$\alpha$ emission line to the host galaxy and not to interaction with a dense CSM. For completeness' sake, we can further consider--if the narrow H$\alpha$ feature is indeed produced by interaction, what CSM density would it imply? Following the method of \cite{Flash2} (which is based on the method of \cite{Ofek2013}), we use the peak flux of the narrow H$\alpha$ emission feature in the $t=2$ day spectrum and constrain the density to be $\rho \lesssim 5\times10^{-15} \mathrm{g \ cm^{-3}}$, which implies a mass loss rate upper limit of $\dot{M} \lesssim 3\times10^{-4} \mathrm{M_\odot \ yr^{-1}}$. This upper limit is consistent with the mass loss rate derived from the P Cygni feature. In conclusion, we assess that the CSM interaction present in SN 2023wdd is not sufficient to warrant the addition of a separate interaction model component to the analytical shock cooling models discussed below.

\subsection{Analytic shock-cooling modeling}

Recently, analytic models of shock cooling have been presented which enable progenitor and explosion parameter estimation from early supernova light curves. As part of our investigation, we estimate progenitor and explosion parameters--particularly the hydrogen envelope mass--using the model of \cite{Sapir2017}. \cite{Sapir2017} extends the model of \cite{Rabinak2011}, which accurately addressed emission generated at $t\lesssim$ a few days post-explosion, when the visible emission is primarily generated close to the surface of the star. \cite{Sapir2017} extends this model to $t\gtrsim$ a few days post-explosion, when the luminosity becomes dependent on the density profile of the material interior to the surface of the star. We follow \cite{Farah2025}, and model the luminosity using the following expression (transformed from \cite{Sapir2017}):
\begin{align}
    L(t) &\sim (1.88-0.147[n-3/2])\times10^{42} \times \nonumber \\
    & \left(\frac{v_{s,8.5} t^2}{f_\rho M \kappa_{0.34}}\right)^{3\times10^{-3}-5.9\times10^{-2}n} \times \nonumber \frac{v_{s, 8.5}^2 R_{13}}{\kappa_{0.34}}\times \nonumber \\
    & \exp\left(-\frac{\left[1.93nt-1.23t\right]}{\left[19.5\kappa_{0.34} M_e v_{s,8.5}^{-1}\right]^{0.5}}\right)^{0.87-4.7\times10^{-2}n} \mathrm{\ erg \ s^{-1}},
    \label{eq:lt_sw17}
\end{align}
where $n$ is the polytropic index ($n=3$ for a radiative envelope, $n=3/2$ for a fully convective envelope) $v_{s,8.5}$ is the velocity of the shock in units of $10^{8.5} \mathrm{\ cm \ s^{-1}}$, $t$ is the phase of the supernova in days, $M$ is the total mass of the star (core $M_c$ plus envelope $M_e$) in solar masses, $\kappa_{0.34}$ is the opacity in units of $0.34 \mathrm{\ cm^2 \ g^{-1}}$, $R_{13}$ is the radius of the extended envelope of the progenitor star $R_e$ in units of $10^{13}$ cm, and
\begin{equation}
f_\rho \sim \begin{cases}\left(M_e / M_c\right)^{0.5} & n=3 / 2 \\ 0.08\left(M_e / M_c\right) & n=3\end{cases}.
\end{equation}
Note that $f_\rho M$ is largely a nuisance parameter and is not expected to be well-constrained. \cite{Sapir2017} also derived a color temperature evolution for the photosphere as a function of time, which we use to fit our multiband datasets. We again refer to the transformed expression used in \cite{Farah2025} and use,
\begin{align}
    T(t) &\sim (2.14-0.06n) \times10^4 \times \nonumber \\
    & \left(\frac{v_{s,8.5}^2 t^2}{f_\rho M \kappa_{0.34}}\right)^{3.8\times10^{-2} - 7.3\times10^{-3}n} \times \frac{R^{0.25}_{13}}{\kappa_{0.34}^{0.25}}t^{-0.5}\mathrm{\ K}.
    \label{sec:fitting:eq:sw17_t}
\end{align}
This model has analytic regimes of validity based on the temperature and transparency timescale of the system. To be consistent with these validity criteria, we only fit data with temperature $\gtrsim 8000$ K and less than the transparency timescale \citep[see][for definition]{Sapir2017} days from explosion, dymamically calculated in the fit depending on the model parameters. See \cite{Farah2025} for further details on our shock cooling emission modeling approach.  

We fit this model using the open-source package \texttt{lightcurve-fitting} \citep{griffin_fitting}, which conditions the \cite{Sapir2017} model on given data using a Markov-Chain Monte Carlo (MCMC) posterior exploration routine. We initialize the fit with 15 walkers and use sufficient steps ($10^3$ steps for burnin in and $\gtrsim 10^4$ steps for posterior exploration) to achieve convergence via the improved Gelman-Rubin statistic and chain autocorrelation length; see \cite{Farah2025} section 5.4 for more details. As is standard, we choose a simple Gaussian likelihood function to compute the likelihood of a particular parameter configuration producing the data. All data are weighted only by their error bars. Parameter estimates and uncertainties are determined by computing the highest posterior density interval around the maximum a posteriori value. This approach is simple but motivated by the marginal parameter posteriors, which are exceedingly well-behaved (i.e., unimodal with minimal skew), with the exception of the nuisance parameter $f_\rho M$.  

For both SN 2023wdd and SN 2022acrv, we varied the polytropic index $n$ between $n=3$ and $n=3/2$ and found that $n=3$ did not provide a satisfactory fit. The fits were unsatisfactory for one or more of the following reasons: (i) the $n=3$ maximum likelihood models described the data significantly more poorly than the $n=3/2$ fits; (ii) the resulting marginal posterior distributions were highly non-Gaussian, with several parameters favoring the edges of the prior, or (iii) the fit favored highly unphysical parameter values, such as envelope masses $\gg 20 \ M_\odot$. The inferior performance of the $n=3$ models suggest that the SN 2023wdd envelope is convective. We also attempted fits both including and excluding non-detections, which did not significantly alter the fit result for either object.

\begin{figure*}
    \centering
    \includegraphics[scale=0.45]{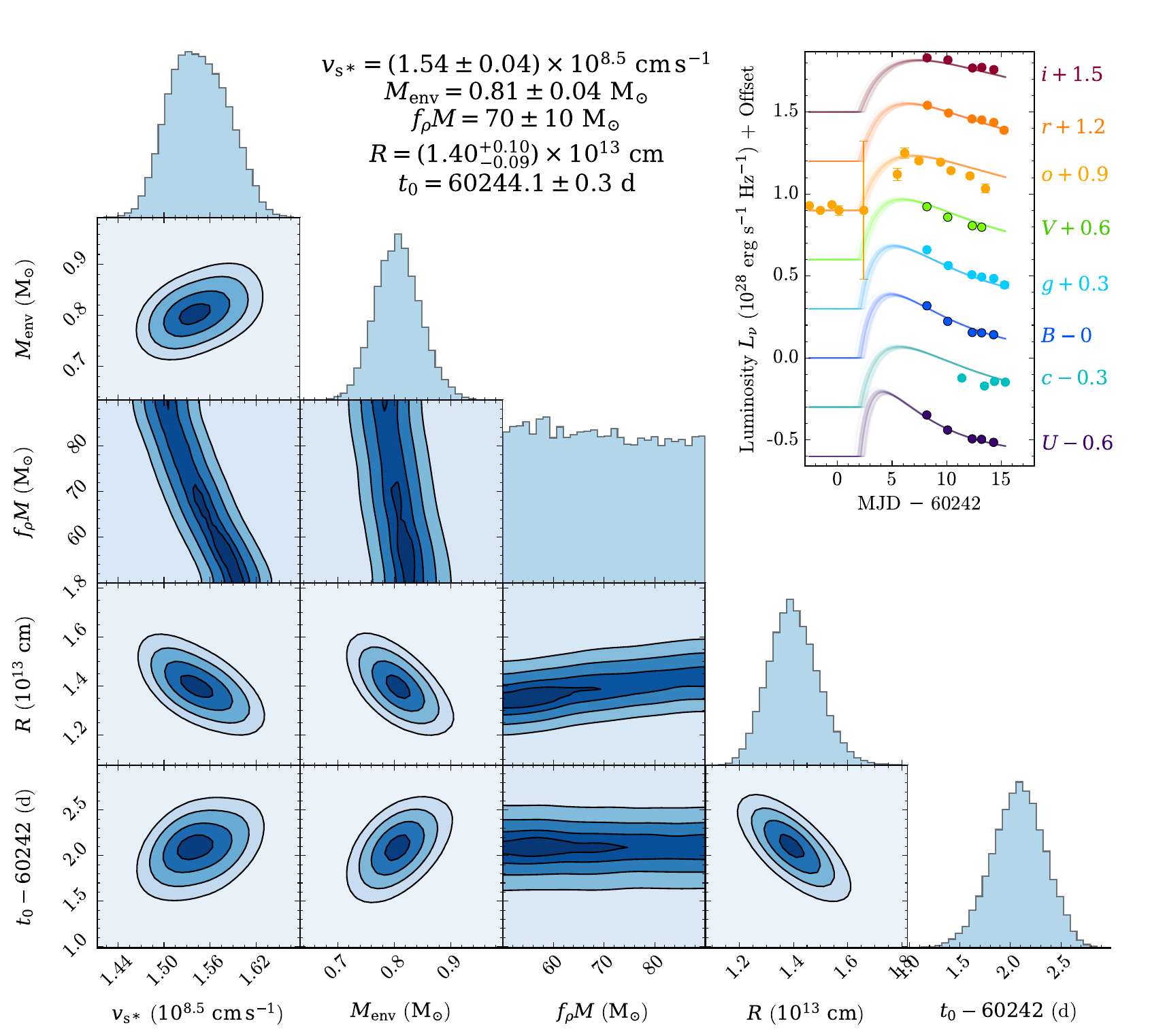}
    \caption{Corner plot and posterior samples for the model of \cite{Sapir2017} with $n=3/2$ applied to the early photometry of SN 2023wdd. We include ATLAS non-detections, which narrowly constrain the explosion epoch. Analytic modeling reports a progenitor radius of $\approx200 \ R_\odot$ and a hydrogen-rich envelope mass of $\approx0.8\pm0.04 \ M_\odot$,  placing SN 2023wdd along the boundary between IIb and short-plateau SNe ($M_{H_\text{env}} \sim 0.91 \ M_\odot$) \citep{Hiramatsu2021}. }
    \label{fig:23wdd_analytic}
\end{figure*}

The fit results for SN 2023wdd are shown in \autoref{fig:23wdd_analytic}. Overall, the model fits well in all bands, with a moderate deviation in ATLAS $o$-band. Best-fit values of the shock velocity ($v_{s\ast} \approx 5\times10^3$ km/s) and progenitor radius ($R\approx 200 \ R_\odot$) are consistent with the broad range of values for other double-peaked objects and indicate a relatively compact progenitor. The best-fit hydrogen-rich envelope mass ($M_{H_\text{env}}\approx 0.8 \pm 0.04 \ M\odot$) is significantly larger than typical double-peaked IIb SNe ($M_{H_\text{env}} \lesssim 0.1 \ M_\odot$), but is consistent with the proposed boundary between IIb and short-plateau SNe ($M_{H_\text{env}} \sim 0.91 \ M_\odot$) \citep{Hiramatsu2021}. 

\label{sub:analytic_shock_cooling_modeling}

\begin{figure*}
    \centering
    \includegraphics[scale=0.4]{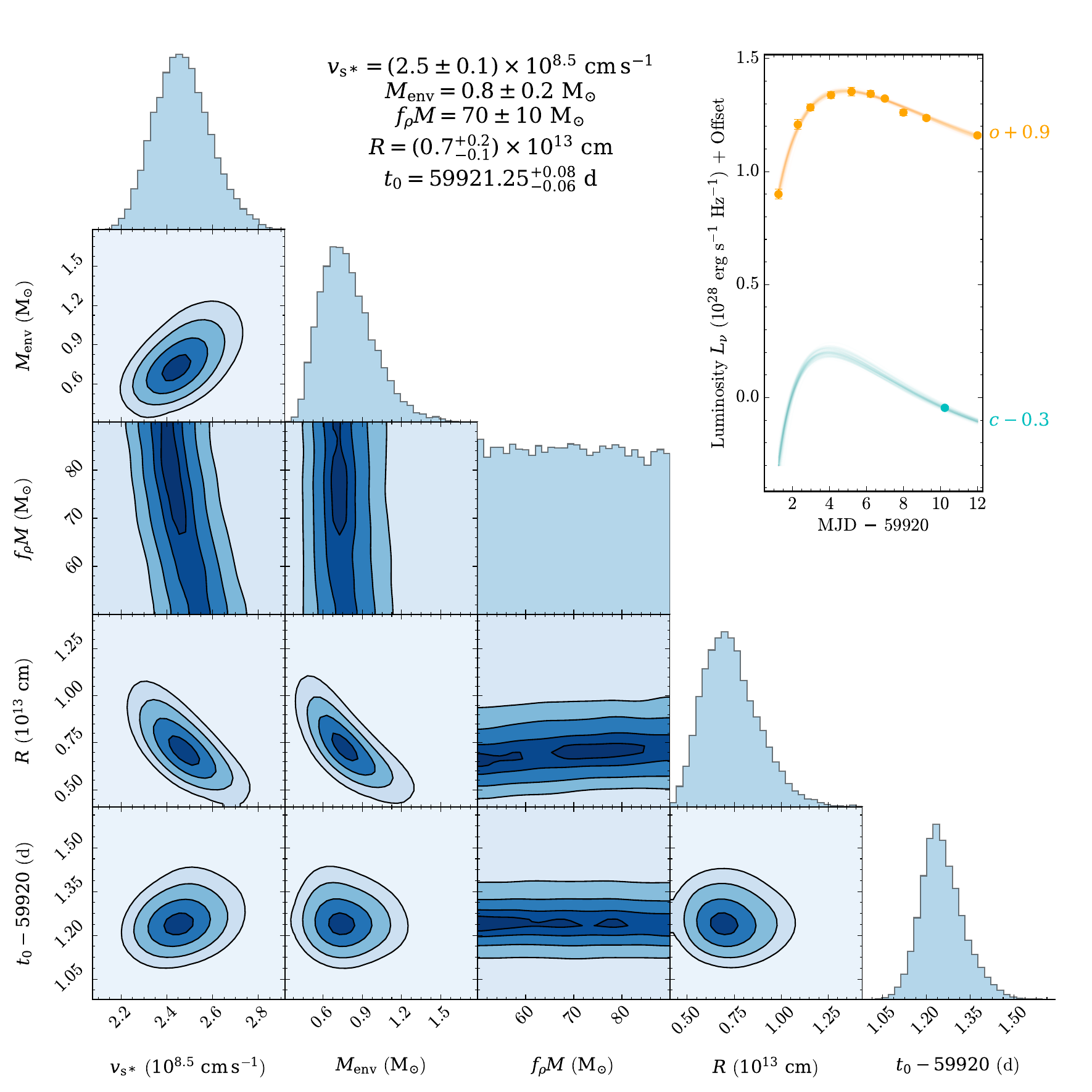}
    \caption{Corner plot and posterior samples for the model of \cite{Sapir2017} with $n=3/2$ applied to the early photometry of SN 2022acrv. Only ATLAS data is available during the shock cooling phase of the light curve. Analytic modeling reports a progenitor radius of $\approx100 \ R_\odot$ and (similar to SN 2023wdd) a hydrogen-rich envelope mass of $\approx0.8\pm0.2 \ M_\odot$,  placing SN 2022acrv along the boundary between IIb and short-plateau SNe ($M_{H_\text{env}} \sim 0.91 \ M_\odot$) \citep{Hiramatsu2021}. The increased uncertainty (an order of magnitude larger than SN 2023wdd) is due to the limited multicolor coverage in this region.}
    \label{fig:22acrv_analytic}
\end{figure*}

The analytic fits for SN 2022acrv (\autoref{fig:22acrv_analytic}) shows significantly more uncertainty, due to the lack of multicolor data for much of the regime of validity. Nevertheless, the results for SN 2022acrv are similar to SN 2023wdd. The shock velocity ($v_{s\ast} \approx 8\times10^3$ km/s) and progenitor radius ($R\approx 100 \ R_\odot$) are again consistent with previous compact double-peak transient progenitors. The constraint on the hydrogen-rich envelope mass is weaker due to the limited data, but we are still able to constrain $M_{H_\text{env}}\approx 0.8\pm0.2 \ M_\odot$, consistent with the proposed IIb-SP SNe boundary.  Thus, the analytic fits for both objects provide evidence that these objects indeed are examples of a transition between the established IIb subclass and the short-plateaus.

\subsection{Numerical modeling via simulation grid}
\label{sub:numerical_modeling_via_simulation_grid}

In addition to analytic models of shock cooling, we model the entire light curve using numerical simulations to identify a progenitor system configuration that best reproduces the observed light curve. \cite{Hiramatsu2021} generated a dense ($N\sim4\times10^4$), discrete grid of numerical simulations varying progenitor zero-age main-sequence masses, mass loss rates, synthesized nickel-56 masses, and explosion energies. We explore this parameter space to identify a maximum likelihood model with corresponding uncertainties.

We begin by computing the bolometric luminosity of each observed light curve, which is then compared directly to simulated light curves. To account for uncertainty in the explosion epoch, each model is allowed an arbitrary shift in time, which is absorbed into the model parameterization. For each model in our discrete parameter grid, we compute a likelihood using a simple Gaussian likelihood function based on the agreement between the model and observed data. To estimate the marginal distributions for each model parameter, we proceed as follows: for each model in the grid, we assign its computed likelihood as a contribution to the corresponding bin of each parameter’s value. Since the parameter grid is not evenly sampled—some parameter values appear more frequently than others—we normalize each bin by the number of models it contains. This yields an average likelihood for each value of each parameter. Finally, we use kernel density estimation \citep{rosenblatt1956remarks,silverman1998density} to convert the discretely sampled bins for each parameter into a continuous marginal probability distribution. These can then be combined to form joint distributions for parameter pairs. Like the analytic modeling, parameter estimates and uncertainties are then determined by computing the highest posterior density interval around the maximum a posteriori value.

\begin{figure*}
    \centering
    \includegraphics[scale=0.67]{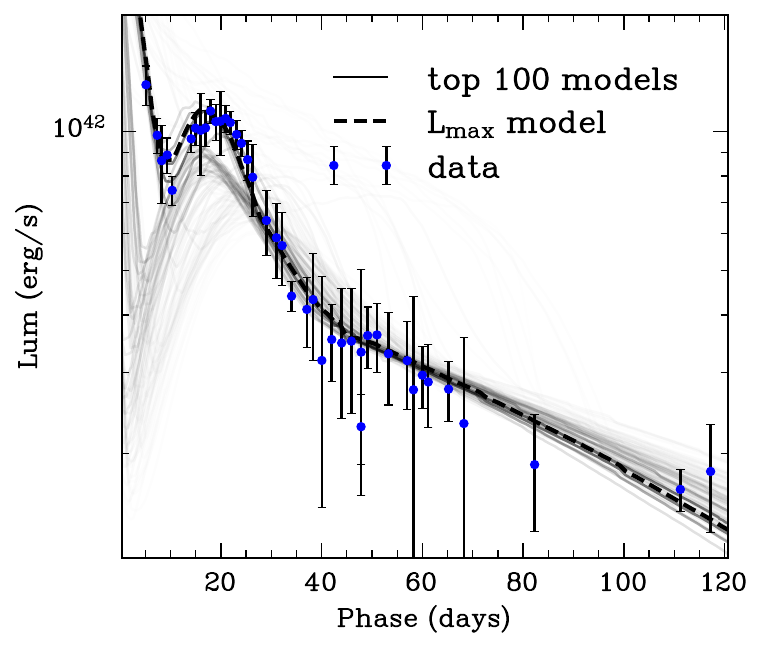}
    \includegraphics[scale=0.5]{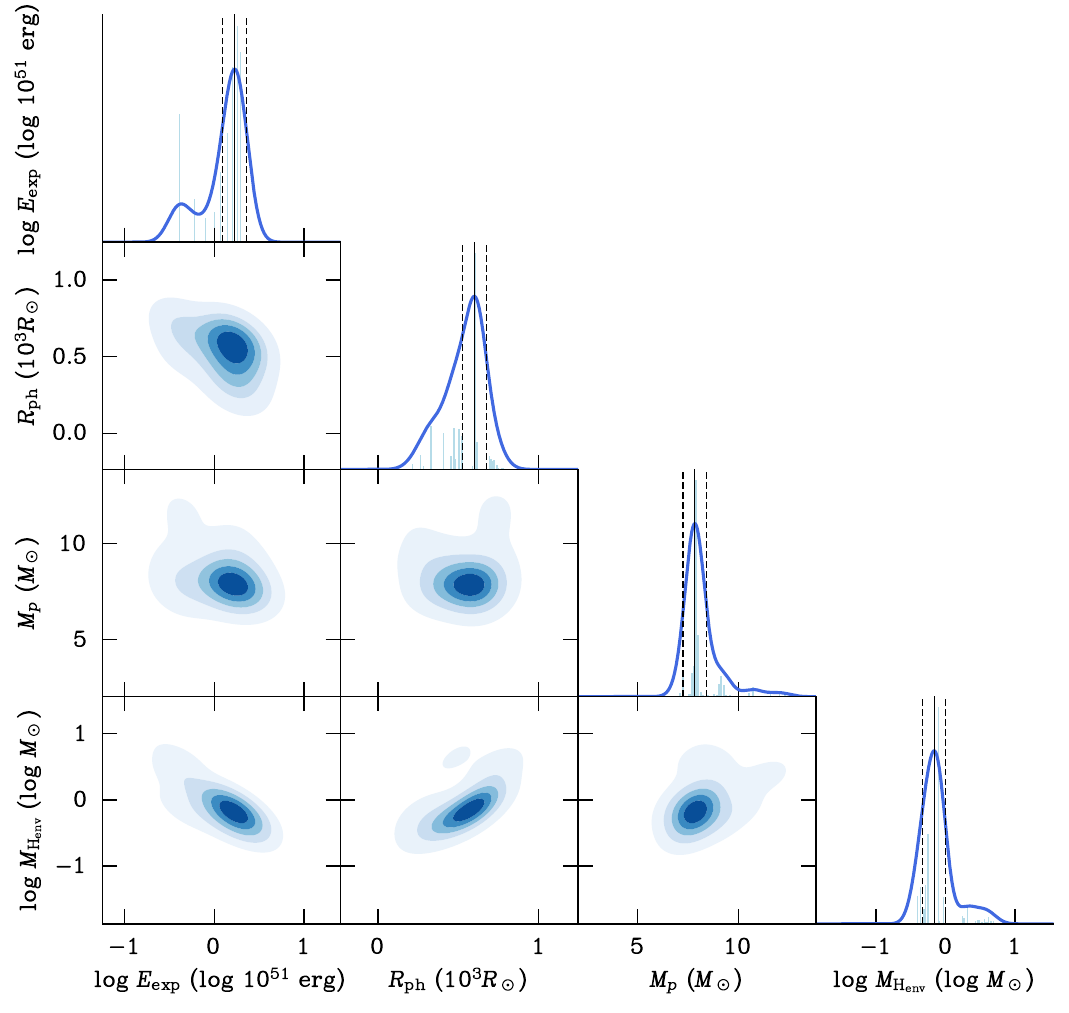}
    \caption{Samples and pseudo-posterior distributions for numerical modeling of SN 2023wdd using the simulation grid of \cite{Hiramatsu2021}. \textit{(left)} Samples from the posterior, ranked by likelihood. The maximum likelihood model is shown as a black dashed line. Sub-maximum likelihood samples are weighted in transparency by likelihood (i.e., darker indicates greater likelihood). \textit{(right)} Psuedo-posterior parameter distributions estimated from the simulation grid. Maximum likelihood parameter estimates are shown in the 1D histograms as solid lines; $\pm1\sigma$ uncertainties are showns as dashed lines. The maximum likelihood simulations have $M_{\text{H}_{\text{env}}}\approx 0.62\pm0.14 \ M_\odot$, consistent up to $1\sigma$ with the analytic modeling, and supporting the characterization of SN 2023wdd as a IIb/short-plateau boundary object.}
    \label{fig:23wdd_numerical}
\end{figure*}

Results of this fitting approach for SN 2023wdd are shown in \autoref{fig:23wdd_numerical}. The top 100 models by likelihood have $M_{\text Ni} = 0.04 \ M_\odot$, consistent with our calculations from the light curve in \autoref{sub:photometric_evolution}. Our maximum likelihood model is consistent with the light curve across all epochs, including the very early shock cooling decline. The best-fit hydrogen envelope mass ($M_{H_\text{env}}\approx0.62\pm0.14 \ M_\odot$) is consistent up to $1\sigma$ with the results from the analytic models, and provides further evidence that SN 2023wdd is a boundary object. However, there is tension (factor of $\approx2.5$) between the progenitor photospheric radius at shock breakout estimated by the model ($R_{\textrm{ph}}\approx 500 R_\odot$) and by the analytic modeling ($R\approx 200 R_\odot$). This discrepancy may be due to (i) simplified assumptions of the stellar density structure present in the analytic models which are more realistically modeled in the numerical grid, or (ii) deviations from reality in the underlying envelope temperature, composition, or density stratification assumed by the analytic model, which can cause order unity shifts in the estimated progenitor radius and shock cooling timescale.

\begin{figure*}
    \centering
    \includegraphics[scale=0.7]{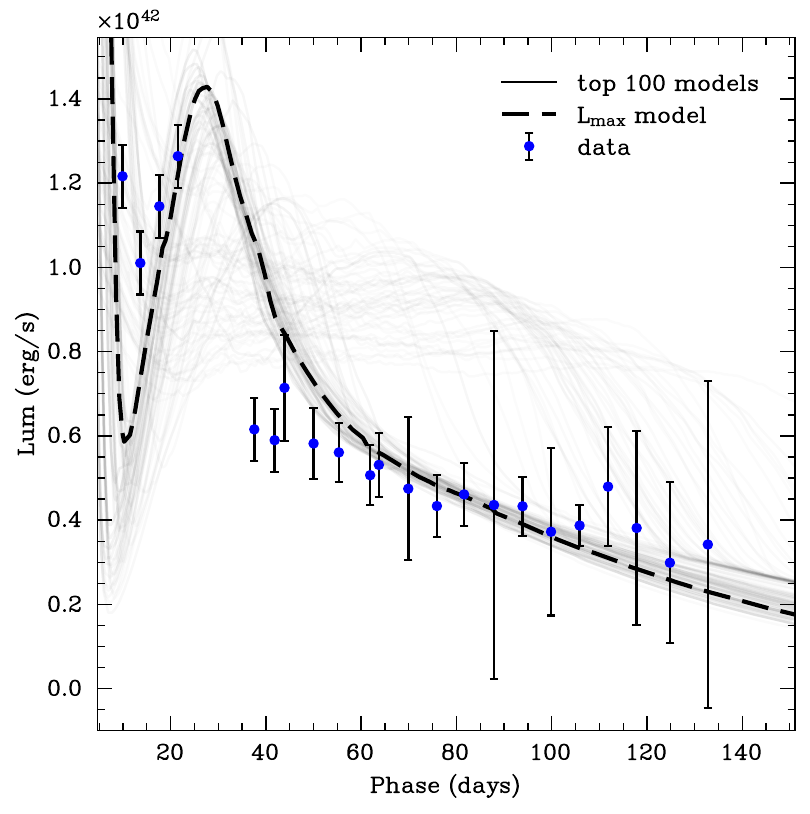}
    \includegraphics[scale=0.4]{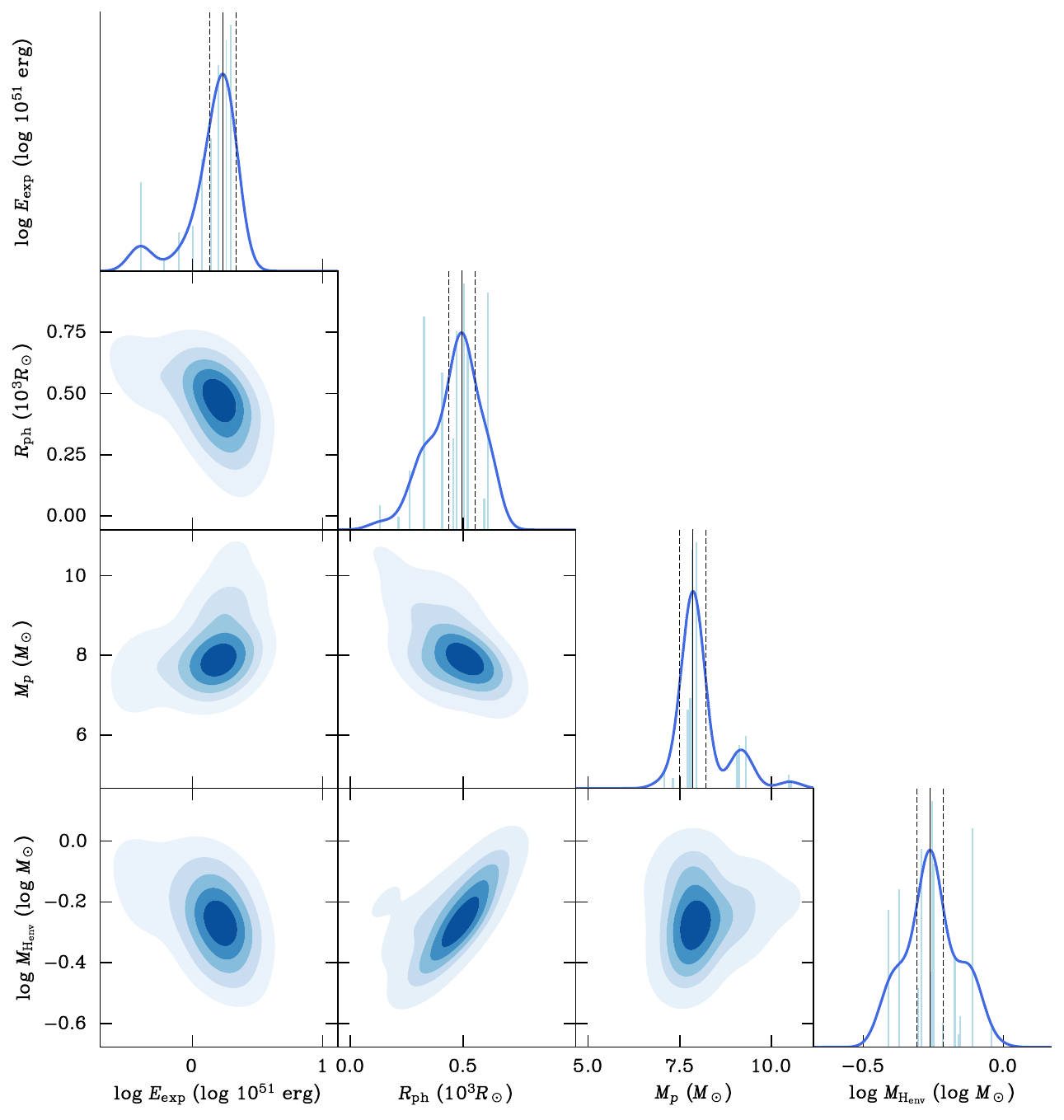}
    \caption{Samples and pseudo-posterior distributions for numerical modeling of SN 2022acrv using the simulation grid of \cite{Hiramatsu2021}. \textit{(left)} Samples from the posterior, ranked by likelihood. The maximum likelihood model is shown as a black dashed line. Sub-maximum likelihood samples are weighted in transparency by likelihood (i.e., darker indicates greater likelihood). \textit{(right)} Psuedo-posterior parameter distributions estimated from the simulation grid. Maximum likelihood parameter estimates are shown in the 1D histograms as solid lines; $\pm1\sigma$ uncertainties are showns as dashed lines. The maximum likelihood simulations have $M_{\text{H}_{\text{env}}}\approx 0.59\pm0.08 \ M_\odot$, consistent up to $1\sigma$ with the analytic modeling, and supporting the characterization of SN 2022acrv as a IIb-SP boundary object. }
    \label{fig:22acrv_numerical}
\end{figure*}

Results of the fitting approach for SN 2022acrv are shown in \autoref{fig:22acrv_numerical}. The top 100 models by likelihood have $M_{\text Ni} = 0.07 \ M_\odot$, consistent with our calculations from the light curve in \autoref{sub:photometric_evolution}. The maximum likelihood model predicts a progenitor radius ($R_{\textrm{ph}}\approx 500 R_\odot$) of similar size to SN 2023wdd and a hydrogen-rich envelope mass ($M_{H_\text{env}}\approx0.59\pm0.08 \ M_\odot$) supportive of the IIb-SP SNe boundary characterization of SN 2022acrv. Consistency of the maximum likelihood model with the data is significantly worse for SN 2022acrv as compared to SN 2023wdd. The model overestimates the object luminosity by $\approx25\%$ on the transition from nickel peak to the radioactive tail, and predicts a dip (between shock cooling decline and nickel peak) $\approx 30\text{--}40\%$ deeper than suggested by the bolometric luminosity. The significant overestimate is limited to a $\sim$few day period immediately after the modeled (but unobserved) peak. Discrepancies between the modeled and observed light curve in this region may be related to parameters of the explosion not explored in our numerical grid, e.g., nickel mixing \citep[see e.g.,][]{Ni56Mixing,Park2024}. Additionally, we note that comparisons in the earlier region of the data may not be reliable, due to the low number of filters available to properly constrain the bolometric properties of the supernova. However, at later times, when significantly more multicolor data is available (LCO+ATLAS), there is more agreement of the maximum likelihood model and the object luminosity. 

Despite a discrepancy between the progenitor radius estimated by the analytic versus numerical modeling channels, all modeling approaches consistently estimate the hydrogen-rich envelope mass of SN 2023wdd and SN 2022acrv to on the boundary of the IIb SNe and short plateau SNe classifications, supporting their characterization as transition objects between these two subclasses. Having established the object characterizations via direct modeling of their progenitor envelopes, we next compare them to other occupants of the Type II/stripped-envelope continuum.

\section{Comparison to other supernovae}
\label{sec:comparisons}

We have established that SN 2023wdd and SN 2022acrv lie upon the boundary between Type IIb and short-plateau SNe. To strengthen this conclusion, we compare these boundary objects to other supernovae with both larger and smaller hydrogen-rich envelope masses. For our sample, we select SN 2016gkg and SN 2022hnt (Type IIb), SN 2023ufx, SN 2006ai, SN 2016egz, and SN 2006Y (short-plateau SNe), and SN 2023ixf, SN 2013ej (Type IIP SNe). These supernova are well-studied, with dense spectrophotometric observations and published constraints on explosion and progenitor parameters. We summarize the objects' properties in \autoref{tab:progenitor_properties}.

\begin{table*}[ht]
\centering
\caption{Hydrogen envelope and plateau properties for various SNe}
\begin{tabular}{lccccccp{5.5cm}}
\bottomrule
Object & Type & $R_p$ ($R_\odot$) & $M_{\text{H}_{\text{env}}}$ ($M_\odot$) & Plateau Length (days) & Reference \\
\hline
SN 2016gkg    & IIb       & $310 \pm 17.9$     & $0.024 \pm 0.003$   & N/A  & \cite{2016gkgYellow,Arcavi2017} \\
\cdashline{1-6}
SN 2022hnt    & IIb       & $70.9 \pm 9$       & $0.068 \pm 0.013$   & N/A  & \cite{Farah2025} \\
\cdashline{1-6}
SN 2023wdd    & boundary  & $560 \pm 80$       & $0.80 \pm 0.04$     & 21   & This work \\
\cdashline{1-6}
SN 2022acrv   & boundary  & $489 \pm 78$       & $0.568 \pm 0.053$   & 23   & This work \\
\cdashline{1-6}
SN 2023ufx    & short-plateau     & $354 \pm 88$       & $1.2 \pm 0.2$       & 24   & \cite{Ravi2025} \\
\cdashline{1-6}
SN 2006Y      & short-plateau     & $480 \pm 100$      & $2.5 \pm 1.2$       & 49   & \cite{Hiramatsu2021} \\
\cdashline{1-6}
SN 2006ai     & short-plateau     & $600 \pm 110$      & $1.7 \pm 1.1$       & 55   & \cite{Hiramatsu2021} \\
\cdashline{1-6}
SN 2016egz    & short-plateau     & $620 \pm 100$      & $2.1 \pm 1.15$      & 74   & \cite{Hiramatsu2021} \\
\cdashline{1-6}
SN 2023ixf    & IIP       & $873 \pm 443$      & $4.5 \pm 0.5$       & 86   & \cite{Hoss2023,Zimmerman2024} \\
\cdashline{1-6}
SN 2013ej     & IIP       & $500 \pm 100$      & $8.55 \pm 2$        & 105  & \cite{Valenti2014} \\
\cdashline{1-6}
\toprule
\end{tabular}
\label{tab:progenitor_properties}
\end{table*}

We begin by comparing to the Type IIb SNe, as both SN 2023wdd and SN 2022acrv were initially classified as Type IIb SNe. Additionally, we compare to the short-plateau SN 2023ufx, described in \cite{Ravi2025}. As we have previously noted, the helium-contaminated H$\alpha$ line which provided the basis for the IIb classification appears to be stronger than expected for a IIb SNe, with weaker contamination than expected. We find a similar behavior in the light curve comparison. Though by eye there appears to be a double peak in both objects, comparison to ``true'' IIb SNe reveals that the peak to be strikingly weak in some bands and arguably nonexistent in others. Particularly in the bluer bands, the extremely shallow dip of the boundary object light curve is more similar to the short plateau light curve behavior (SN 2023ufx) than the IIb light curves. SN 2023ufx is especially interesting for comparison, as with a modeled hydrogen-rich envelope mass of $M_{\textrm{H}_{\textrm{env}}} \approx 1.2 \pm 0.2 \ M_\odot$, it lies along (more specifically, just above) the IIb/short-plateau boundary ($M_{\textrm{H}_{\textrm{env}}}\approx 0.91 \ M_\odot$) as well. 

\begin{figure*}
    \centering
    \includegraphics[scale=0.41]{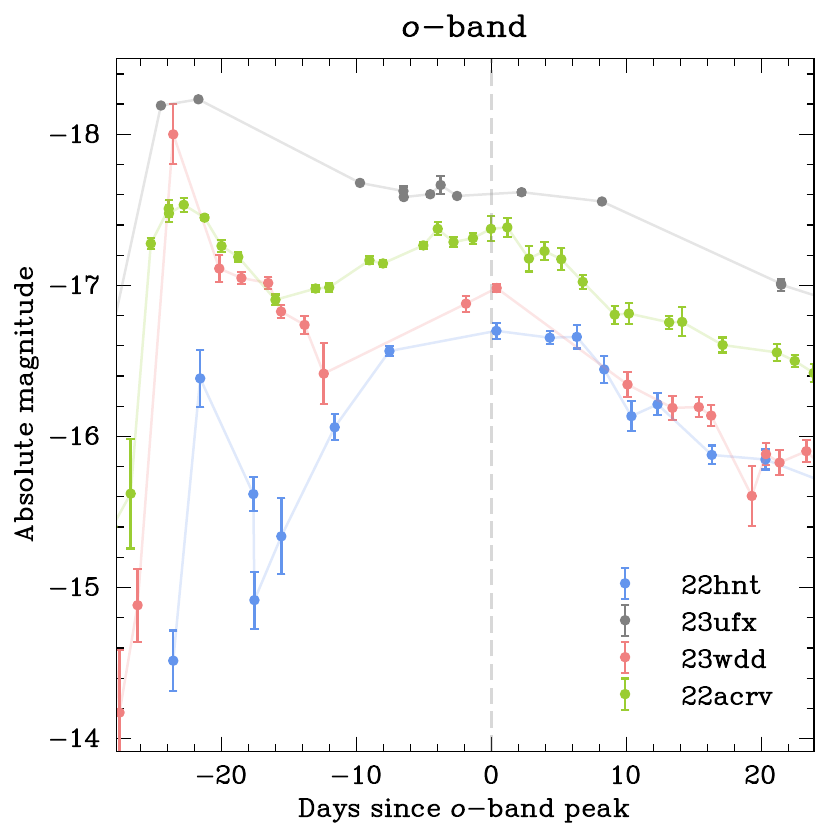}
    \includegraphics[scale=0.41]{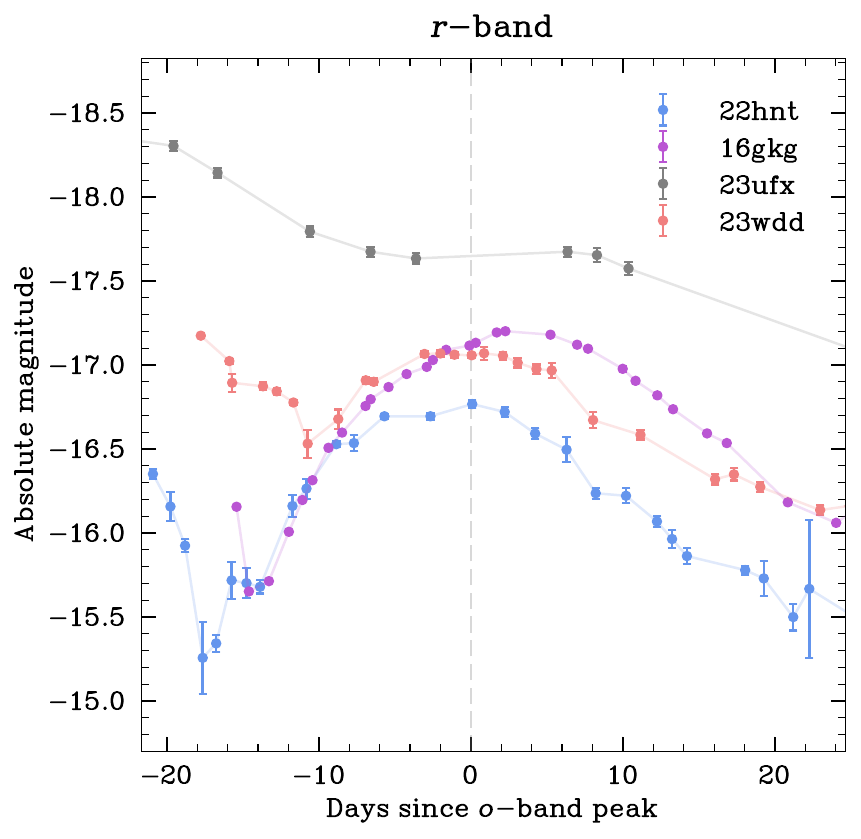}
    \includegraphics[scale=0.41]{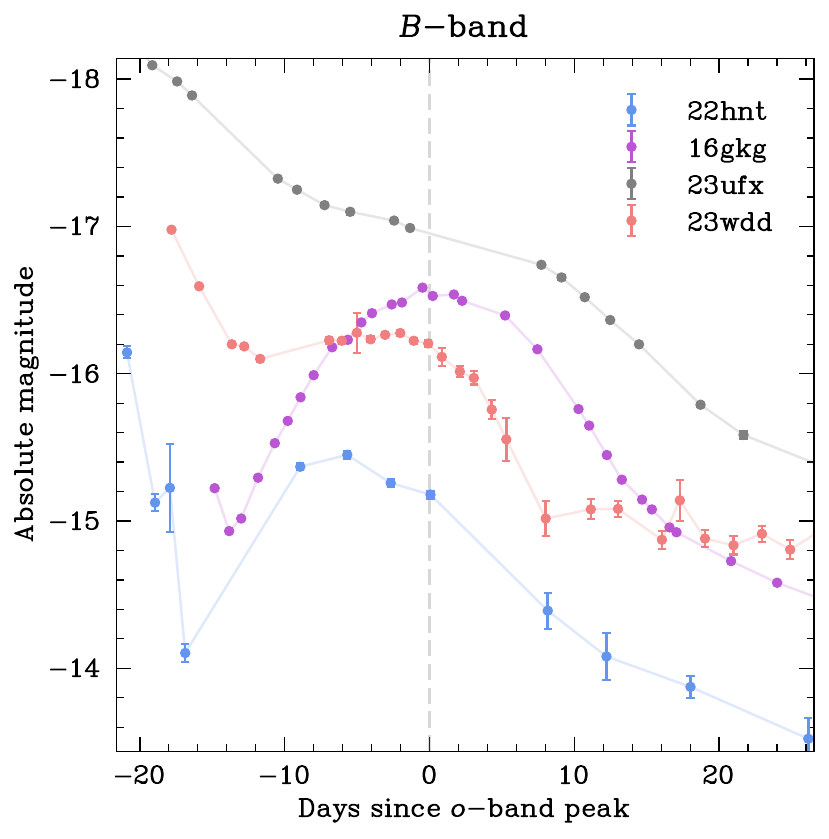}
    \caption{Comparison of filter photometry for the early evolution of SN 2023wdd, SN 2022acrv, and two established IIb SNe (SN 2022hnt and SN 2016gkg). Despite appearing to be double-peaked by eye, SN 2023wdd and SN 2022acrv have a significantly lower dip-to-peak distance than conventional IIb SNe, owing to the more hydrogen-rich envelope. This pattern becomes more apparent in $r$-band and $B$-band. Particularly in $B$-band, the light curve of SN 2023wdd looks almost plateau-like by comparison to SN 2016gkg and SN 2022hnt. Such behavior is reasonable for an object on the boundary between double-peaked IIb SNe and objects with short plateaus.}
    \label{fig:filter_comparison}
\end{figure*}

The $o$-, $r$, and $B$- band light curves of SN 2022acrv and SN 2023wdd are compared to SN 2016gkg and SN 2022hnt in \autoref{fig:filter_comparison}. We characterize all light curves based on a ``dip-to-peak'' distance in magnitudes, following the ``lightning bolt'' model of \cite{crawford2025peakyfinderscharacterizingdoublepeaked}. That work characterized this phase of the evolution of Type IIb SNe using the \texttt{m3} parameter, which describes the slope of the second rise to nickel peak. Our ``dip-to-peak'' distance is the product of the \texttt{m3} parameter and the rise time to the second peak. We directly the ``dip-to-peak'' distance by taking the nickel peak absolute magnitude and subtracting it from the minimum absolute magnitude achieved during the dip demarcating the double peak. Beginning with $o$-band, we observe that SN 2022hnt experiences an absolute ``dip-to-peak'' distance of $\approx1.8$ mag. By contrast, SN 2023wdd and SN 2022acrv execute almost identical ``dip-to-peak'' distances of $\approx0.6$ and $\approx0.5$ mag, respectively. This is a factor of $\sim3$ in magnitude space less prominent of a dip in $o$-band for the boundary objects as compared to the Type IIb SNe. 

Next, we compare the behavior in $r$-band in a similar way to $o$-band. While SN 2023wdd executes a ``dip-to-peak'' distance of $\approx0.7$ mag (comparable to $o$-band), the IIb SNe SN 2016gkg and SN 2022hnt display ``dip-to-peak'' distances of $\approx 1.5$ mag. This is a factor of $\sim2$ in magnitude space more prominent of a dip for the Type IIb SNe as compared to SN 2023wdd. 

Finally, we compare the evolution in $B$-band, where the difference in behavior is stark. As in all other bands, the Type IIb SNe execute prominent dips, with ``dip-to-peak'' distances of $\approx 1.5$ mag, similar to $r$- and $o$- bands. However, in $B$-band, SN 2023wdd hardly executes a dip at all, with all brightnesses during the dip consistent with a plateau at absolute mag $\approx16.2$. In the most charitable interpretation of the data, the maximum possible ``dip-to-peak'' distance in this band is $\lesssim 0.1$ mag. 

The weak double-peak behavior--in contrast to the more heavily hydrogen-rich Type IIb SNe--can be explained by the boundary objects possessing more hydrogen in their outer envelopes than the Type IIb SNe prior to explosion. This explanation is consistent with theoretical predictions from \cite{Hiramatsu2021}, as well as the modeled hydrogen-rich envelope masses, which found $M_{\textrm{H}_{\textrm{env}}} \lesssim 0.1 \ M_\odot$ for the Type IIb SNe and $M_{\textrm{H}_{\textrm{env}}} \approx 0.8 \ M_\odot$ for the boundary objects.  

\begin{figure*}
    \centering
    \includegraphics[scale=0.5]{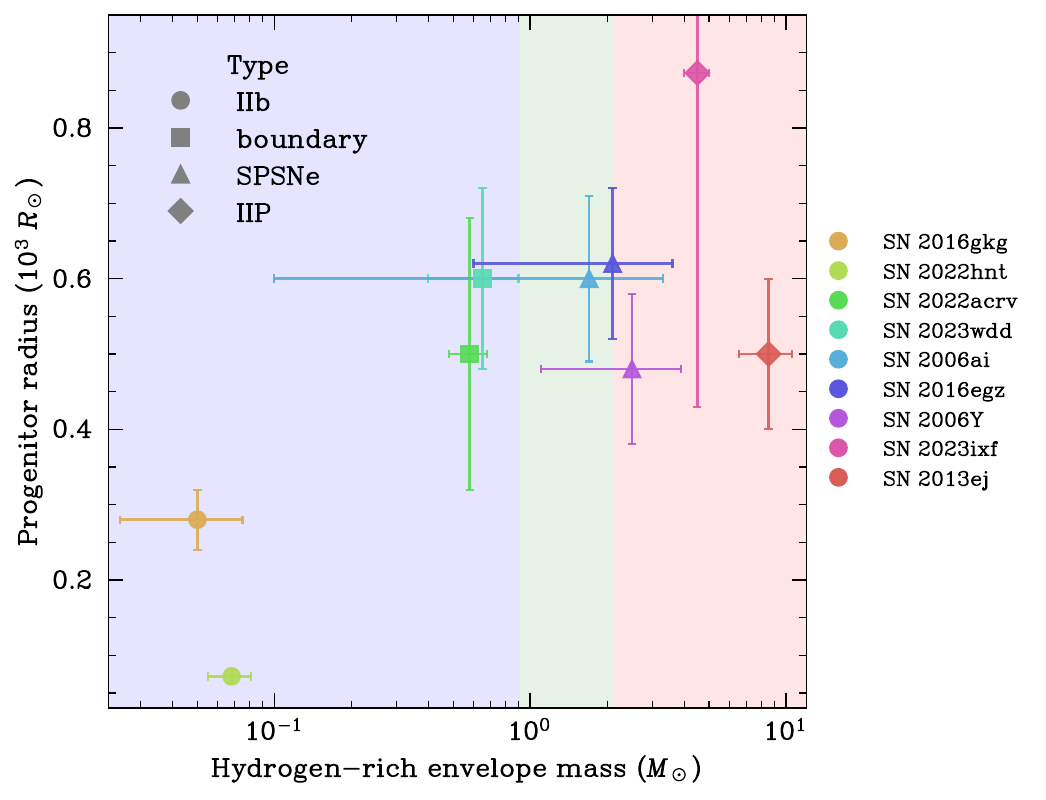}
    \includegraphics[scale=0.5]{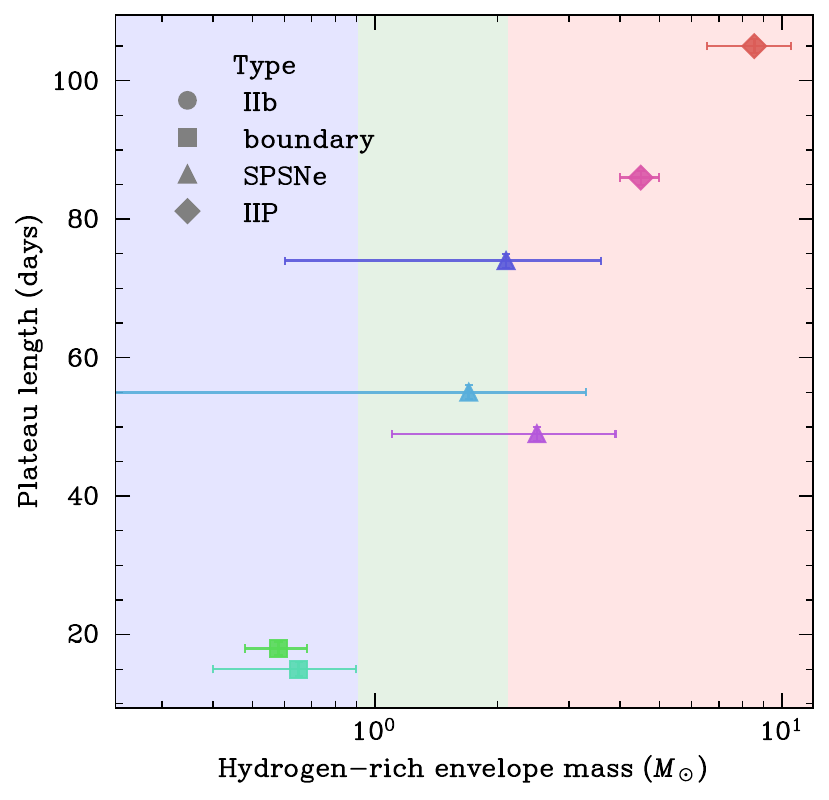}
    \caption{Comparison of progenitor properties across the spectrum of stripped envelope objects. In addition to our boundary objects, we examine established IIb SNe (SN 2022hnt and SN 2016gkg), candidate SPSNe (SN 2006ai, SN 2016egz, SN 2006Y), and established Type IIP SNe (SN 2023ixf and SN 2013ej). (See \autoref{tab:progenitor_properties} for object property references.)  Regions of the plot are shaded based on the predicted hydrogen-rich envelope mass ranges for IIb ($M_{\text{H}_{\text{env}}}\lesssim0.91 \ M_\odot$), short-plateau ($0.91 \ M_\odot \lesssim M_{\text{H}_{\text{env}}}\lesssim2.1 \ M_\odot$), and IIP/L ($M_{\text{H}_{\text{env}}} \gtrsim 2.12 \ M_\odot$) SNe proposed in \cite{Hiramatsu2021}. The marker for each supernova identifies its established or proposed classification, based on spectroscopic or photometric evolution. \textit{(left)} Comparison of hydrogen-rich envelope mass and progenitor radius for all objects. There is remarkable consistency between object classification, estimated hydrogen envelope mass, and the subtype separation proposed by \cite{Hiramatsu2021} (colored regions). \textit{(right)} Comparison of hydrogen-rich envelope mass and plateau length for all objects. The IIb SNe are not shown, as they do not have a plateau. For the boundary objects, we interpret the weak dip as a plateau for comparison purposes. There is a clear trend between plateau length and hydrogen-rich envelope mass, consistent with expectations from analytic \citep[][shown as red spread]{Popov1993} and numerical theory \citep{Hiramatsu2021}. The theoretical prediction of \cite{Popov1993} is plotted here based on the spread of inferred explosion energies, hydrogen-rich envelope masses, and progenitor radii for the objects in the sample. }
    \label{fig:supernova_comparison}
\end{figure*}

In addition to light curve behavior, we compare inferred progenitor properties to other objects along the Type II/stripped-envelope continuum. Across the supernovae in our sample, we evaluate hydrogen-rich envelope mass, progenitor radius, and approximate plateau length and compare to theoretical predictions. A visualization of this comparison is shown in \autoref{fig:supernova_comparison}. Proposed classifications (IIb, short plateau, IIP SNe) based on the hydrogen-rich envelope mass ranges presented in \cite{Hiramatsu2021} are shown as shaded regions in the figure. We find that the placement of SN 2023wdd and SN 2022acrv along the Type II/stripped-envelope continuum based on their inferred hydrogen-rich envelope mass is consistent with our assessment from spectrophotometric data that they lie upon the boundary between IIb and short-plateau SNe. Further, if the weak dip feature in SN 2022acrv and SN 2023wdd is instead treated as a ``plateau'', we find it is in strong agreement with the lower-$M_{\textrm{H}_{\textrm{env}}}$-shorter-plateau trend observed among other core-collapse supernovae, as well as with predictions from analytic \citep{Popov1993} and numerical theory \citep{Hiramatsu2021}. In fact, given the unusual part-dip, part-plateau behavior of the light curve, SN 2023wdd in particular may represent a sort of ``missing link'', potentially the highest $M_{\textrm{H}_{\textrm{env}}}$ Type IIb SN (or the shortest short-plateau SN) possible.  

There are several factors contributing to why such boundary objects have not been better characterized in past studies of hydrogen-rich supernovae. First, Type IIb supernovae are a relatively rare outcome of the core collapse of massive stars. \cite{Li2011} found that Type IIb SNe make up $\approx11\%$ of their volume-limited sample, which more recent studies have corroborated \citep{Shivers2017}. Even among the already rare IIb subclass, double-peaked IIb SNe (which provide in their double peak a direct probe of $M_{\text{H}_{\text{env}}}$) occur $\lesssim48\%$ of the time \citep{ayala2025earlylightcurveexcess}. The low volume of IIb SNe generally available for study makes refining the boundaries of this subclass challenging; however, future surveys \citep[e.g., Legacy Survey of Space and Time;][]{LSST} are expected to greatly increase the rate of Type IIb supernovae discovery by orders of magnitude. Additionally, these boundary objects may be challenging to identify as they are often classified as Type IIb SNe (due to the presence of helium in the spectra), which typically have a larger dip-to-peak distance (or steeper \texttt{m3}) than objects on the IIb/short-plateau boundary. Indeed, multiple studies have found significant diversity in the early excess emission behavior of objects classified as Type IIb SNe \citep{crawford2025peakyfinderscharacterizingdoublepeaked,ayala2025earlylightcurveexcess}, but in this view, boundary objects such as SN 2023wdd and SN 2022acrv are seen as outliers in the Type IIb subclass, and not as probing a smooth continuum of hydrogen envelope stripping. 
\section{Conclusions}
\label{sec:conclusions}

Our observations and analysis of SN 2023wdd and SN 2022acrv reveal two objects on the boundary of the distinct domains of Type IIb and short-plateau supernovae (short-plateau). Despite sharing similarities with double-peaked IIb SNe, both show comparatively weaker early declines (or very short plateaus), and spectroscopic features in which H$\alpha$ remains significant even as modest He I contamination appears. By fitting both analytic shock-cooling models and the numerical simulation grid of \cite{Hiramatsu2021}, we estimate hydrogen envelope masses of $\sim0.6\text{--}0.8 \ M_\odot$, placing these events above the typical hydrogen mass range for double-peaked IIb SNe ($\lesssim 0.1 \ M_\odot$) yet just below that of the proposed short-plateau subclass ($\gtrsim 0.91 \ M_\odot$). This quantitative constraint, along with their spectral and photometric evolution, provides supporting evidence for a continuum of hydrogen envelope stripping in massive stars before core collapse.

Notably, both supernovae show ``weaker'' second peaks than canonical double-peaked IIb SNe. In SN 2023wdd, the transition from the shock-cooling decline to the radioactive peak is especially subtle in bluer bands, approaching a short-plateau appearance more similar to the behavior of SNe such as SN 2023ufx rather than the pronounced dip observed in well-studied IIb SNe like SN 2016gkg or SN 2022hnt. Such behavior theoretically follows from a more hydrogen-rich outer envelope in the progenitor prior to explosion. Although the light-curves of SN 2023wdd and SN 2022acrv could be characterized as borderline short-plateau SNe, their helium contamination and partial envelope stripping align them more closely with IIb-like progenitors, and thus place them along the boundary between Type IIb SNe and short-plateau SNe predicted by recent theoretical work.

The findings here underscore the importance of obtaining well-sampled, multi-band photometry and high-cadence spectroscopy in the first few weeks following explosion. Such observations are key to distinguishing subtle transitions between ``hydrogen-poor'' and ``hydrogen-rich'' envelope configurations, especially in this intermediate regime. As new large-scale transient surveys deliver hundreds of core-collapse supernovae each year, the detection of additional boundary objects will illuminate the physical processes enabling fine-grained variations in progenitor hydrogen-rich envelope mass. Ultimately, building a robust sample of these transitional events will help unify our understanding of how mass loss shapes the diversity of hydrogen-rich explosions.

\acknowledgements{We thank Grant 1, 2, and 3.}

\bibliography{ref}


\end{document}